\documentclass[aps,prb,twocolumn,amsmath,amssymb,showpacs,floatfix,superscriptaddress]{revtex4}
\usepackage{amsmath}
\usepackage{graphicx}
\usepackage{epsfig}
\begin{document}
\newcommand{\dg}{^{\dagger }}
\newlength{\upit}\upit=0.1truein
\newcommand{\raiser}[1]{\raisebox{\upit}{#1}}
\newlength{\bxwidth}\bxwidth=1.5 truein
\newcommand\frm[1]{\epsfig{file=#1,width=\bxwidth}}
\def\fig#1#2{\includegraphics[height=#1]{#2}}
\def\figx#1#2{\includegraphics[width=#1]{#2}}
\newlength{\figwidth}
\newcommand{\fg}[3]
{
\begin{figure}[ht]
\[
\includegraphics[width=\figwidth]{#1}
\]
\vspace*{-4mm}
\caption{\label{#2}
\small
#3
}
\end{figure}}
\newcommand{\mat}[4]{\left[
\begin{array}{cc}
#1 & #2 \cr #3 & #4
\end{array} \right]
}
\newcommand{\cmat}[4]{\left (
\begin{array}{cc}
#1 & #2 \cr #3 & #4
\end{array} \right)
}
%\psdraft
\title
{\bf Conductance of a spin-1 quantum dot: the two-stage Kondo effect}

\author{A. Posazhennikova}
\email[Email: ]{anna@tfp.physik.uni-karlsruhe.de}

\author{B. Bayani}

\affiliation{
Institut f\"ur Theoretische Festk\"orperphysik and DFG-Center of Functional Nanostructures,
Universit\"at Karlsruhe,
 D-76128 Karlsruhe, Germany}

\author{P. Coleman}
\affiliation{Center for Materials Theory,
Department of Physics and Astronomy, Rutgers University, Piscataway, NJ
08854, USA}

\preprint{version of \today}

%\date{}

\begin{abstract}
We discuss the physics of a 
of a spin-1 quantum dot, coupled to two
metallic leads and develop a simple model for the temperature dependence of
its conductance. Such quantum dots are described by a two-channel Kondo
model with asymmetric coupling constants and the spin screening of the dot 
by the leads is  expected to proceed via a two-stage process. 
When the Kondo
temperatures of each channel are widely separated, on cooling,
the dot passes through a broad cross-over regime dominated by
underscreened Kondo physics. A singular, or non-fermi liquid correction to
the conductance develops in this regime.  
At the lowest temperatures,  destructive interference between resonant scattering in both channels
leads to the eventual suppression of the
conductance of the dot. 
We develop a model to describe the
growth, and ultimate suppression of the conductance in the two channel
Kondo model as it is screened successively by its two channels.  Our
model is based upon large-N approximation in which the localized
spin degrees of freedom are described using the Schwinger boson
formalism.
\end{abstract}
\maketitle

%\newpage
\section{Introduction}

The Kondo effect arises from the resonant spin-flip scattering of
electrons off localized magnetic impurities, which manifest themselves
through the anomalous transport and
thermodynamic properties of dilute magnetic alloys
\cite{Hewson}. A modern context for the physics of local moments is
found within quantum dots \cite{Glazman01} in where the Kondo effect
is manifested as a zero-bias anomaly of the differential
conductance. The high degree of tunability in quantum dots allows for
realization of a number of interesting variants of the Kondo effect,
including both the ``over-screened'' and ``under-screened'' Kondo
models.  For 
example, the Kondo effect has been observed in 
spin-1 quantum dots containing an even number of electrons, by tuning
the dot to a singlet-triplet degeneracy point by
applying a magnetic field \cite{sasaki00,kogan03}.

The physics of a single spin $1/2$ magnetic impurity, with no orbital degrees of
freedom is described by the canonical one-channel Kondo model, where 
the loan spin is fully screened by the conduction electrons at
low temperatures, forming a paramagnetic ground-state with excitations
described by Landau Fermi liquid theory. 
Nozi\`eres and Blandin pointed out 
\cite{Nozieres80} that in more realistic magnetic ions the orbital
structure of the local impurity electron needs to be taken into account
($l\neq 0$), giving rise  to a much richer class of spin-screening
phenomena.   
For a spin $S$ with unquenched orbital angular momentum
$l$, there are $n_{l}=2l+1$ screening channels.  
Depending on whether $n_{l}$ is greater than, equal to, or smaller than
the number $n_{e}=2S$ of Hund's coupled electrons in the local moment,
the spin is said to be overscreened ($2l+1>2S$), fully screened ($2l+1=2S$) or
underscreened ($2l+1< 2S$). In both the underscreened and overscreened cases,
the excitations of the ground-state are no longer described by a Fermi liquid,
and a variety of exotic non-Fermi liquid (NFL) phenomena can develop,
as summarized in  Table I.
Unfortunately, most transition metal ions are described by the
fully screened case, where $2S=2l+1$, so the
underscreened and overscreened Kondo effect is not readily observed in
bulk materials. For this reason, quantum dots provide
an important alternative milieu for the study of unconventional Kondo screening
phenomena. 

Most recent interest has focused on the case of 
the overscreened multichannel Kondo model. From detailed
Bethe ansatz\cite{Andrei84} and conformal field theory solutions\cite{affleck91}
the low-energy physics of the 
overscreened $S=1/2 $ Kondo effect is known to involve a non-trivial fixed
point, with singular power-law behavior of the specific
heat $C\sim T^{\frac{4}{n_l+2}}$ and magnetic susceptibility. 
(Logarithmic corrections to the
specific heat appear in the two-channel case).  
A possible realization of this kind of behavior 
in the heavy electron compound $UBe_{13}$ has been proposed 
by Cox, where quadrupole  fluctuations of  singlet uranium ions
may assume the role of spin fluctuations in the conventional Kondo
effect\cite{Cox87,Andrei04}. 

In point of fact, 
a spin-1/2 quantum dot coupled to two leads does not provide a
realization of the two-channel Kondo effect, because the localized 
spin couples to a specific  combination of the lead electrons forming
a one-channel Kondo model\cite{Glazman88}. 
A proposal to overcome this problem through the coupling of a quantum
dot to a strongly interacting ``quantum box'' has been advanced 
by Oreg and Goldhaber-Gordon\cite{Oreg03}, and the first, preliminary indications of the 
the two-channel Kondo effect have been recently reported\cite{Rau}.

\begin{table*}
\caption{Ground state properties of the Kondo model with impurity spin
$S$ and number of channels $n_l$.
 $C_v$ is the specific heat, and $\chi$ the magnetic
susceptibility. 
In the over-screened Kondo
effect, electron-scattering remains inelastic in the ground-state.}\vskip 0.2truein
\label{tab_ideal}
\begin{tabular}{|c||c|c|c|c|c|}
\hline
Kondo effect & Examples & GS & Phase shift $\delta (\epsilon)$& $C_v$ & $\chi$\\
\hline
\hline
Fully Screened & $S=1/2; n_l=1$ & Fermi & $\frac{\pi}{2}+\alpha \epsilon$ & $C_v\sim T$ & $ \sim \frac{1}{T_{K}}$\\
 &  $S=1; n_l=2$ & Liquid & & &   \\
\hline
&&&&&\\
Under Screened  & $S=1; n_l=1$ & Singular & $\frac{\pi}{2}+\frac{\alpha}{\ln(T_K/\epsilon)}$   & $\sim \frac{1}{T \ln^{4} (T/T_{K})} $ & $\sim \frac{1}{B \ln^{2} (B/T_{K})}$  \\
&&Fermi Liquid&&&\\
\hline
Over Screened & $S=1/2; n_l=2$ & Non &  inelastic  & $\sim T \ln T$ & $\sim \ln T$\\
 &$S=1; n_l=3$ & Fermi Liquid & scattering & $\sim T^{4/5}$ & $\sim T^{-1/5}$ \\
\hline
\hline
\end{tabular}

\end{table*}

By contrast, realization of the underscreened Kondo effect requires a
quantum dot with a higher spin $S>1/2$.  In most quantum dots,
electrons fill the dot in accordance with the Pauli principle, forming
a singlet state when the number of electrons is even, and for the most
part, zero-bias anomalies are observed in dots
with odd numbers of electrons
\cite{Gordon98,Cronenwett98}.  
A sequence of more recent experiments have shown that zero-bias
anomalies associated with a Kondo effect can also occur in quantum
dots with even occupancies, where Hund's coupling between the
electrons lead to novel degeneracies, either through the formation
of higher spin states, or through the accidental degeneracy of singlet
and triplet states.  The observation of a zero-bias anomaly in even
integer quantum dots was first reported by Schmid et al.\cite{schmid00}
who tentatively identified the phenomenon with a triplet ground state
of the quantum dot. Sasaki {\it et al}\cite{sasaki00} later discovered a
zero-bias anomaly in even electron quantum dots that are tuned to the
degeneracy point between singlet and triplet states. Most recently,
Kogan {\it et al}\cite{kogan03} have shown that the singlet - triplet
excitation energy in lateral quantum dots can be tuned by the gate
voltage, explicitly demonstrating that the zero bias anomaly develops
once the triplet state drops below the singlet configuration.

Pustilnik and Glazman \cite{Pustilnik01} have shown 
that a spin-$1$ quantum dot is associated with {\it two} screening channels which fully screen the
local moment at the lowest temperatures.  The two channels arise because the two electrons in a  spin-1
dot occupy two separate orbitals which couple differently to the leads. At low temperatures the 
conductance of the Fermi liquid that emerges is governed by the
interference  between the two channels 
\[
G = 2\frac{e^2}{h} \sin ^2 ( \delta_1 - \delta_2)
\]
where $\delta_1$ and $\delta_2$ are the scattering phase shifts of the two
screening channels \cite{Pustilnik01}. According to this line of reasoning, 
the development of a unitary phase shift in each channel, $\delta_1=\delta_2=
\pi/2$ leads to 
a complete suppression of the zero bias anomaly in a
triplet quantum dot \cite{zarand04}.

The two separate screening channels of a spin-1 quantum dot are associated with 
two different 
antiferromagnetic coupling constants $J_{\lambda}$ ($\lambda=1,2$). Since the Kondo temperature
depends exponentially on the coupling constants, $T_{K\lambda}=D\sqrt{J_{\lambda}\rho}\exp({-\frac{1}{J_{\lambda}\rho}})$ ($\rho$ is the density of states and $D$ is the band width), a rather modest
difference in coupling constants can produce Kondo temperatures
separated by decades of temperature, $T_{K1}>>T_{K2}$. Over the
intervening temperature range, defined as
$ln(T_{K1})\gg ln(T)\gg ln(T_{K2})$,  the physics of the dot is
expected to be dominated by the underscreened fixed point, 
or one-channel spin-1 Kondo model. 
In the underscreened spin-1 Kondo, the
residual spin-1/2 moment is {\it ferromagnetically} coupled to leads,
with a coupling that scales asymptotically  towards zero
\cite{Nozieres80}. Recent studies of this problem have argued that the
ground-state which develops is a ``singular Fermi liquid'', in which
the electrons are elastically scattered with unitary phase shift, but
in which the logarithmically decaying coupling to the partially
screened local moment leads to a singular energy dependence in the
scattering phase shift and a divergence in the resulting quasiparticle
density of states\cite{pepin,pankaj,indranil}. This singular behavior,
has a number of observable consequences for the conductance which if
confirmed, will provide a first experimental realization of singular
Fermi liquid behavior of an underscreened local moment.

In this paper we present an analysis of the transport properties of a
spin-1 quantum dot in two cases: first for one screening channel and
then for two active screening channels.  The initial part of our paper
provides a  detailed treatment of the singular
conductance properties previously predicted by us in a short paper on
this topic\cite{Anna05}. The second part of the paper extends this
earlier work to take account of the second-channel screening processes
and the interference it gives rise to. One of the main results of this
new analysis, is the development for an approximate expression for the
temperature dependence of the linear conductance. We find that the
linear conductance can be divided into a coherent part, and an
incoherent part,
\[
G = G_{coh}+ G_{inc}
\]
where
\begin{equation}\label{}
G_{coh}=\frac{Ne^{2}}{h}(2\alpha\beta )^2\int d\omega
\left(- \frac{df (\omega)}{d\omega} \right)
\nonumber \\
\vert \pi\rho ( T_{1} (\omega)- T_{2} (\omega))\vert^{2}
\end{equation}
defines the coherent conductance through the two-channel and
\begin{eqnarray}\label{l}
G_{inc}&=&\frac{Ne^{2}}{h}(2\alpha\beta )^2 
\int d\omega
\left(- \frac{df (\omega)}{d\omega} \right)\cr
&\times& \pi\rho[\tau^{-1}_{1} (\omega)+\tau^{-1}_{2} (\omega)]
\end{eqnarray}
the additional contribution due to incoherent transport, where
``$T_{1}$'' and ``$T_{2}$'' refer to the scattering t-matrices in the 
two screening channels of the quantum dot, $\alpha $ and $\beta $
define the relative amplitudes of channel $1$ in the left and right
channels respectively 
and 
\begin{equation}\label{}
\tau^{-1}_{\lambda}= {\rm Im}T_{\lambda} (\omega)- \pi \rho \vert
T_{\lambda} (\omega)\vert^{2}.  \qquad (\lambda=1,2)
\end{equation}
are the inelastic  scattering rates in each channel, defined by the
deviation of the imaginary part of the t-matrix from the value
expected from the optical theorem.

\section{Construction of the Hamiltonian}

Our initial model Hamiltonian for a spin-1 quantum
dot coupled to two leads can be divided up into a Hamiltonian for the
leads ($H_{L}$), the quantum dot ($H_{D}$) and the coupling between the leads and the dot  ($H_{C}$) as follows
\begin{equation}\label{}
H= H_{L}+ H_{C}+H_{D}
\end{equation}
where 
\begin{equation}\label{}
H_{L}=\sum_{\stackrel{k\sigma}{\gamma=L,R}}\epsilon_k c_{k\gamma\sigma}\dg c_{k\gamma\sigma}
\end{equation}
describes the electrons in the left and right leads,
\begin{equation}\label{hybrid}
H_{C}=t_1 \sum_k (\psi\dg_{k\sigma}d_{1\sigma}+{\rm H.c}.)+t_2\sum_k ( \varphi\dg _{k\sigma}d_{2\sigma}+{\rm H.c.}) ,
\end{equation}
describes the hybridization in two channels between the lead electrons and the quantum
dot through the contacts. Here $d\dg _{1\sigma }$ and $d\dg_{2\sigma
}$ create electrons in two orthogonal states of the quantum dot and
\begin{eqnarray}
\psi\dg _{k\sigma} &=&\alpha c\dg _{kL\sigma}+\beta c\dg _{kR\sigma} \nonumber \\
\varphi_{k\sigma} &=&-\beta c\dg _{kL\sigma}+\alpha c\dg _{kR\sigma},
\label{two_channel}
\end{eqnarray}
create the linear combinations of conduction electrons that hybridize
with these two orthogonal scattering channels.  The Hamiltonian of the spin-1
quantum dot can be written
\begin{eqnarray}\label{l}
H_{D}&=& -E_1n_{d1}-E_{2}n_{d2}+
U_{1} n_{d1\uparrow}n_{d1\downarrow }+U_{2} n_{d2\uparrow}n_{d2\downarrow }\cr
&+& U_{12}n_{d1}n_{d2} -J [\vec{S}_{1}\cdot \vec{S}_{2}].
\end{eqnarray}
Here  $-E_{\lambda}$ ($\lambda=1,2$)  are the
energies of the two one-particle states  and 
$n_{d\lambda}=\sum_{\lambda}d\dg_{\lambda \sigma
}d_{\lambda\sigma }$ ($\lambda = 1,2$) are the occupancy of the two
channels. $U_{1}$, $U_{2}$ and
$U_{12}$ are the intra and inter-channel Coulomb interactions and
lastly $\vec{ S}_{\lambda}= \frac{1}{2}d\dg_{\lambda \alpha }\vec{\sigma}_{\alpha
\beta }d_{\lambda\beta }$ are the spin operators for the two channels
and $J$ is the (ferromagnetic) Hund's exchange coupling between
the spins.

In general, we are interested in the case where the one-particle
energies $-E_{1}$ and $-E_{2}$
are {\sl negative}, but $U_{1}$, $U_{2}$ 
are large quantities that restrict double occupancy of either
d-state. Finally, the Hund's interaction $J$ is assumed to be large
restricting the doubly occupied state to be a spin-1 triplet state.
In this situation, the quantum dot is restricted to the states
\begin{eqnarray}\label{l}
\vert d^{0}\rangle&\equiv &\vert 0\rangle , \cr\cr
\vert d^{1}:1 \sigma \rangle& =& d\dg_{1\sigma }\vert d^{0}\rangle,  \qquad 
\vert d^{1}:2 \sigma \rangle = d\dg_{2\sigma }\vert d^{0}\rangle, \cr\cr
\vert d^{2}:M\rangle &\equiv&
\left\{\begin{array}{lc}
d\dg_{1\uparrow}d\dg_{2\uparrow}\vert d^{0}\rangle,&(M=1)\cr
\frac{1}{\sqrt{2}} (
d\dg_{1\downarrow}d\dg_{2\uparrow}+d\dg_{1\uparrow}d\dg_{2\downarrow})\vert d^{0}\rangle ,&(M=0)\cr
d\dg_{1\downarrow}d\dg_{2\downarrow}\vert d^{0}\rangle ,&(M=-1)
\end{array}
 \right.
\end{eqnarray}
The energy of the $d^{2}$ triplet states is
$E_{1}+E_{2}-J/4+U_{12}$ (see Fig.1). 

\begin{figure}[t]
\includegraphics[width=0.45\textwidth]{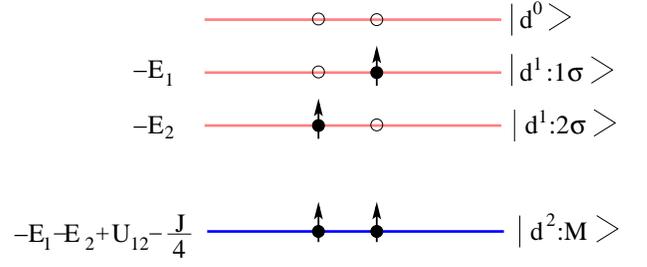}
\caption{Energy diagram of a spin-1 quantum dot with strong Hund's coupling.}
\label{cond_temp}
\end{figure}

We can project the Hamiltonian onto this Hilbert space by combining the d-electron operators
with a generalized Gutzwiller projection operator as follows, 
\begin{equation}\label{}
d\dg_{\lambda\sigma }
\rightarrow 
P_{T}d\dg_{\lambda\sigma } \equiv  X\dg_{\lambda\sigma }
\end{equation}
where
\begin{equation}\label{}
P_{T}= P_{G1}P_{G2}
\left(1+\vec{S}_{1}\cdot\vec{S}_2 - \frac{n_{d1}n_{d2}}{4}
 \right)
\end{equation}
and
\begin{equation}\label{}
P_{G\lambda}=(1- n_{d\lambda\uparrow}n_{d\lambda\downarrow }), \ \ (\lambda=1,2)
\end{equation}
is the Gutzwiller projector for channel one and two.
Here, we have used Hubbard's  ``X'' notation to describe the projected
d-electron fields. 
With this notation, we can write our projected Hamiltonian as
\begin{eqnarray}\label{themodel}
H&=&\sum_{k\sigma}\epsilon_k
\psi_{k\sigma}^+\psi_{k\sigma}+\sum_{k\sigma}\epsilon_k
\varphi_{k\sigma}^+\varphi_{k\sigma} \nonumber \\&+&t_1 \sum_k
(\psi_{k\sigma}^+ X_{1\sigma}+h.c.)+t_2\sum_k ( \varphi_{k\sigma}^+
X_{2\sigma}+h.c.) \nonumber \\
&-&E_1\sum_{\sigma}X_{1\sigma}^+X_{1\sigma}-E_2\sum_{\sigma}X_{2\sigma}^+X_{2\sigma}\cr
&+& \left(U_{12}-\frac{J}{4} \right)n_{d1}n_{d2}
\label{And_2chx} 
\end{eqnarray}
The main interactions in this Hamiltonian are hidden in the
constraints. In the Kondo limit, where this Hamiltonian is dominated
by excitations between the $d^{2}$ and $d^{1}$ state, the final
interaction term has the effect of replacing the excitation energy
\[
-E_{\lambda}\rightarrow E (d^{2})-E (d^{1}) =- E_{\lambda}-\frac{J}{4}+U_{12}
\]
With this observation, we shall eliminate the final interaction from
the Hamiltonian, absorbing its effect into the above redefinition of $E_{\lambda}$. (Another
way is to set $U_{12}=J/4$).

In the extreme Kondo limit where the dot is predominantly in the
$\vert d^{2}\rangle $
state, the $d^{2}\leftrightarrow d^{1}$ charge fluctuations can be
eliminated from this Hamiltonian via a Schrieffer Wolff
transformation to give
\begin{eqnarray}\label{l}
H=H_{L}+J_{1}\sum_{k,k'\alpha \beta }\psi \dg_{k\alpha }\vec{ \sigma}_{\alpha \beta }\psi_{k'\beta}\cdot \vec{S} 
\nonumber \\
+J_{2}\sum_{k,k'\alpha \beta }\phi \dg_{k\alpha }\vec{ \sigma}_{\alpha \beta }\phi_{k'\beta}\cdot \vec{S} 
\end{eqnarray}
where
\begin{equation}\label{}
J_{1}= \frac{( t_{1})^{2}}{\frac{J}{4}+ E_{1}-U_{12}}, 
\quad
J_{2}= \frac{( t_{2})^{2}}{\frac{J}{4}+ E_{2}-U_{12}}, 
\end{equation}
are the two antiferromagnetic coupling constants, corresponding to two orthogonal channels.

\section{Schwinger Boson representation}\label{}

A very convenient way to treat the Gutzwiller operators in this
Hamiltonian is to use
a combination of slave fermions and Schwinger bosons. If we describe
the  empty state $\vert d^{0}\rangle= \chi \dg_{2}\chi \dg_{1}\vert 0\rangle  $ as a
pair of two slave fermions,
then the singly occupied states
are written
\begin{eqnarray}\label{l}
\vert d^{1}:1\sigma \rangle \equiv b\dg_{\sigma}\chi_{1}\vert
d^{0}\rangle &=& -b\dg_{\sigma }\chi\dg _{2}\vert 0\rangle ,\cr
\vert d^{1}:2\sigma \rangle \equiv b\dg_{\sigma}\chi_{2}\vert
d^{0}\rangle&=& b\dg_{\sigma }\chi\dg _{1}\vert 0\rangle ,
\end{eqnarray}
while the doubly occupied states are given by
\[
\vert d^{2}:M\rangle 
= \left\{ 
\frac{1}{\sqrt{2}}
( b\dg_{\uparrow })^{2}\vert 0\rangle ,\
b\dg_{\uparrow}b\dg_{\downarrow }\vert 0\rangle , \ \frac{1}{\sqrt{2}} (b\dg_{\downarrow })^{2}\vert 0\rangle ,
 \right\}
\]
In this representation,  
the Hubbard operators are written
\begin{align}\label{hubbardops}
X\dg _{1\sigma } = \frac{1}{\sqrt{n_{b}}}
b\dg_{\sigma }\chi_{1}, 
\qquad 
X\dg _{2\sigma } = \frac{1}{\sqrt{n_{b}}}b\dg_{\sigma }\chi_{2}, 
\end{align}
The prefactors $\frac{1}{\sqrt{n_{b}}}$ ($n_b$ is the number of Schwinger bosons) are needed to guarantee the
normalization of the wavefunction.     Thus
\begin{equation}\label{}
\vert d^{2}:M=1\rangle = X\dg_{1\uparrow}X\dg_{2\uparrow }\vert
0\rangle = \frac{1}{\sqrt{2}} (b\dg_{\uparrow})^{2}\vert 0\rangle 
\end{equation}
which is the right normalization. Similarly 
\begin{eqnarray}\label{l}
P_{T}d\dg_{1\uparrow}d\dg_{2\downarrow}\vert d^{0}\rangle &=&
X\dg_{1\uparrow}X\dg_{2\downarrow }\vert d^{0}\rangle \cr
&=& \frac{1}{\sqrt{2}}b\dg_{\uparrow}b\dg _{\downarrow }\vert d^{0}\rangle 
= \frac{1}{\sqrt{2}}\vert d^{2}:m=0\rangle.
\end{eqnarray}
which consistently normalizes the matrix element 
\begin{equation}\label{}
<d^{2}:m=0\vert X\dg_{1\uparrow }\vert d^{1}:1\downarrow \rangle =\frac{1}{\sqrt{2}}.
\end{equation}
The physical Hilbert space
is defined by those states where 
\begin{equation}\label{}
n_{b}+n_{\chi 1}+ n_{\chi 2}= 2S
\end{equation}

\subsection{Large-N expansion}

To
develop a controlled many body treatment of \eqref{And_2chx} we shall
employ a large-$N$ expansion, extending the number of spin components $\sigma $
from two to $N$. 
For the work in this paper, we are interested in the approach to the Kondo
limit of this problem, where the amplitude of the valence fluctuations
are small, and in this limit, we shall replace $\sqrt{n_{b}}\rightarrow \sqrt{2S}$ in
(\ref{hubbardops}) the above expression. With this device, 
\[
t_{\lambda}X\dg_{\lambda \sigma } \longrightarrow
\frac{\tilde{t}_{\lambda}}{\sqrt{N}}b\dg_{\sigma }\chi_{\lambda}, \qquad (\lambda=1,2)
\]
where $N$ is the spin degeneracy and $\tilde{t}_{\lambda}= t_{\lambda}/\sqrt{2S/N}$.
The model Hamiltonian is then 
\begin{eqnarray}\label{newmodel}
H&=&\sum_{k\sigma}\epsilon_k
\psi_{k\sigma}^+\psi_{k\sigma}+\sum_{k\sigma}\epsilon_k
\varphi_{k\sigma}^+\varphi_{k\sigma} \nonumber \\&+& 
\frac{{\tilde{t}_{1}}}{\sqrt{N}}
(\psi_{k\sigma}^+ \chi \dg_{1}b_{\sigma }
+h.c.)+\frac{{\tilde{t}_{2}}}{\sqrt{N}}
\sum_k ( \varphi_{k\sigma}^+
\chi \dg_{2}b_{\sigma }+h.c.) \nonumber \\
&+&E_1 \chi \dg _{1}\chi _{1}+E_2 \chi \dg _{2}\chi_{2}.
\label{And_2ch} 
\end{eqnarray}
where it is understood that as $N$ becomes large, $\tilde{t}_{1,2}$ is
kept fixed.
There is one additional trick required to preserve a finite scattering
phase shift as $N\rightarrow \infty$.  
For this task, 
we introduce $K=kN$ bosonic ``replicas'', where $k$ is maintained
fixed as $N\rightarrow \infty$. With this device, in strong coupling
$K$ bosons bind into the Kondo singlet and 
we can produce a large $N$ mean field theory in which the scattering phase shift is $\delta=\pi k$, with the qualitatively correct logarithmic energy dependences \cite{indranil}. The Hamiltonian used in the large $N$ expansion is then
\begin{eqnarray}\label{main}
H&=&\sum_{k\sigma}\epsilon_k
\psi_{k\sigma}^+\psi_{k\sigma}+\sum_{k\sigma}\epsilon_k
\varphi_{k\sigma}^+\varphi_{k\sigma} \nonumber \\&+& \frac{\tilde{t}_{1}}{\sqrt{N}}
\sum_{k\sigma \mu}(\psi_{k\sigma}^+ \chi \dg_{1\mu}b_{\sigma \mu}
+h.c.) \nonumber \\ &+&\frac{\tilde{t}_{2}}{\sqrt{N}}\sum_{k\sigma \mu} ( \varphi_{k\sigma}^+
\chi \dg_{2\mu}b_{\sigma \mu}+h.c.) \nonumber \\
&+&\sum_{\mu}\left(E_1 \chi \dg _{1\mu}\chi _{1\mu}+E_2 \chi \dg
_{2\mu}\chi_{2\mu}\right)\cr
&+& \lambda (n_{b}+n_{\chi 1}+ n_{\chi 2}- 2SK)
\end{eqnarray}
where the sum over $\mu$ runs from $1$ to $kN$, while the last term
imposes the generalized constraint 
\begin{equation}\label{newconstraint}
n_{b}+n_{\chi 1}+ n_{\chi 2}= 2SK.
\end{equation}
Notice how the large $N$ model Hamiltonian \eqref{main} is obtained by
replacing
\begin{equation}\label{}
t_{\lambda}\longrightarrow \frac{\tilde{t}_{\lambda}}{\sqrt{N}},
\qquad  X\dg _{\lambda \sigma }\rightarrow  \sum_{\mu=1,K}b\dg_{\sigma
\mu}\chi_{\lambda \mu}.
\end{equation}
in \eqref{themodel}.
Section \ref{sec_US} 
examines in the underscreened one-channel
regime, where $\tilde{t}_{2}=0$. Section \ref{sec_twoch} proceeds with
a discussion of the full Hamiltonian.

\section{One-channel spin-1 quantum dot: underscreened Kondo effect} \label{sec_US}

\subsection{The singular Fermi liquid}

In this section we discuss the limit where 
the Kondo temperature of the second-channel $T_{K2}$ is negligibly
small, so that the physics of the quantum dot is governed by an
underscreened Kondo effect. We  present a large $N$ approximation to
the physics of this underscreened limit. 

The Hamiltonian that governs this behavior is derived from the Hamiltonian \eqref{main} with $\tilde t_2=0$ and $E_2=0$. 
Much is known about the equilibrium physics of this model. 
At low temperatures, the spin is partially screened from spin $S$ to
spin $S- (1/2)$. The residual moment is ferromagnetically coupled to the
conduction sea, with a residual coupling that slowly flows to weak
coupling according to 
\[
J\rho  (\Lambda) = -\frac{1}{\ln (
\frac{T_{K1}}{\Lambda})
} +
O\left(\frac{1}{\ln^{2}(\frac{T_{K1}}{\Lambda})
 } \right)
\]
where $\Lambda\sim max (T,\mu_{B}B)$ is the characteristic cut-off
energy scale, provided in equilibrium, by the temperature or magnetic
field. 
At low energies and temperatures, the partially
screened magnetic moment scatters electrons elastically, with a unitary phase
shift, however the coupling to the residual spin $( S- \frac{1}{2})$ gives rise
to a singular energy dependence of the scattering phase shift. The low
energy scattering phase shift can be deduced from scaling theory 
to have the  asymptotic form 
\[
\delta(\omega) = \frac{\pi}{2}\pm \pi \rho J (\omega)= 
\frac{\pi}{2} \left(1 \pm \frac{( S-\frac{1}{2})}{\ln
(T_{K1}/\omega)} \right).
\]
for the ``up'' and ``down'' spin channels. 
The logarithmic term on the right hand side is produced by the residual
coupling between the electrons and the partially screened moment. 
While the electrons at the Fermi energy  scatter elastically off the local moment with unitary
scattering phase shift, as in  a Fermi liquid, 
the logarithmically singular dependence of the phase shift leads to a divergent
density of states, $N (\omega)\sim \frac{1}{\pi}\frac{d \delta
(\omega)}{d\omega}\sim \frac{1}{\vert \omega\vert }$, which means that
we can not associate this state with a bona-fide Landau Fermi
liquid. For this reason, the ground-state of the underscreened Kondo
model has recently been called a ``singular Fermi
liquid''\cite{pankaj}. 

These singular features of the underscreened Kondo effect are expected 
to manifest themselves in the properties of a triplet quantum
dot, in the range where $T_{K2}<<T<<T_{K1}$.
In this regime, the physics is dominated by the underscreened fixed
point.
For example, as a function of magnetic field, we expect the
conductance to follow the simple relation
\[
G (B) = \frac{dI }{dV}= \frac{e^{2}}{h}
\sin^2
\delta (B)
\]
so at low fields $T_{K1}<<\mu_B B<< T_{K2}$, the differential conductance will have the form 
\[
G \sim 
\frac{e^{2}}{h}\left(1 - \frac{\pi^{2}}{16}
\frac{1}
{ln^{2} \left(\frac{T_{K1}}{\mu_B B} \right)} 
\right)
\] 
Notice that the field derivative of the conductance diverges as $1/B$
\[
\frac{dG}{dB}= \frac{1}{B}
\frac{e^{2}}{h}\left(
\frac{\pi^{2}}
{8  ln^{3} 
\left[\frac{T_{K1}}{ \mu_B B} \right]
}
\right)
\]
at low fields. 
The prediction of the finite temperature, and finite voltage
conductance can not be made exactly, however we expect the above form
to hold, for the differential conductance at finite temperature or
voltage, with an appropriate replacement of cut-offs, namely 
\[
G (V,T) 
\sim 
\frac{e^{2}}{h}\left(1 - \frac{\pi^{2}}{16}
\frac{1}
{ln^{2} \left(\frac{T_{K1}}{{\rm max} (T, eV)} \right)} 
\right)
\]
so that over the temperature range dominated by the underscreened Kondo
fixed point, the conductance will be a monotonically increasing function
of decreasing temperature and voltage. Over this range, 
$dG/dT\sim 1/T$ and
$dG/dV\sim 1/V$ will be divergent functions of temperature and voltage
respectively.

\subsection{Dyson equations and one-channel t-matrix}

We now analyze the low energy behavior of the t-matrix in the large-N
approach. Our development begins with a derivation 
of the Dyson equations for the Green's function of the  
conduction $\psi_{k\sigma}$-fermions ($G$) that are coupled to the
dot and the Green's function of auxiliary $\chi-$holons ($J$) that
describe the partial screening. 
In this large-N approach, we will show that the Green's function of the Schwinger bosons $B$ remains unrenormalized.  

In general, we will be dealing with a voltage-biased quantum dot. The
Hamiltonian for the leads is given by 
\begin{eqnarray}
H_0&=&\sum_{k\gamma\sigma} (\epsilon_k - \mu_{\gamma})c\dg
_{k\gamma\sigma}c_{k\gamma\sigma} \nonumber , \qquad (\gamma=R,L)
\end{eqnarray}
where 
$\mu_L=+\frac{eV}{2}$ and $\mu_R=-\frac{eV}{2}$. 
The Green's function of the conduction electrons is given by 
\begin{equation}
g_\gamma (t-t') =-i\sum_k \langle T_{c_K} c_{k\gamma\sigma} (t) c\dg _{k\gamma\sigma} (t')  \rangle
\end{equation}
where $c_K$ is the Keldysh contour. Following the standard procedure,
we write this Green's function using the Larkin-Ovchinnikov
representation, in terms of the advanced ($g^{A}$), retarded ($g^{R}$) and Keldysh Green's
functions $g^{K}$as follows
\[
\hat g_{\gamma} = \left[\begin{array}{cc}
g_{\gamma}^{R} & g_{\gamma}^{K}\cr 0 & g_{\gamma}^{A}
\end{array} \right]
= \left[\begin{array}{cc}
-i \pi\rho & - i2\pi \rho (1 - 2 f_{\gamma})\cr 0 &  i \pi \rho 
\end{array} \right]
\]
where $\rho $ is the electron density of states and
\[
f_{\gamma}(\epsilon)\equiv  f (\epsilon- \mu_{\gamma}),\qquad (\gamma
= R,L)
\]
describes the Fermi function in the left and right-hand leads.
The complete Green's function describing the two leads is then a
four-dimensional, block-diagonal matrix
\begin{equation}
\hat g_{lead}=\left(
\begin{array}{cc}
\hat g_{L} & 0 \\
0 & \hat g_{R}
\end{array}
\right)
\label{g_lead}
\end{equation}
To obtain the corresponding Green's function in the channel basis, we
carry out a unitary transformation, writing 
\[
\hat  g_{ch} = \left[\begin{array}{cc} \hat g_{\psi  \psi } & \hat g_{\psi  \varphi}\cr
\hat g_{\varphi\psi} & \hat g_{\varphi \varphi}\end{array} \right]=R \hat g_{lead} R^{-1}
\]
where
\begin{equation}
R=\left( 
\begin{array}{cc}
\alpha
 & \beta\\
-\beta & \alpha 
\end{array}
\right).
\end{equation}
transforms between the two bases. The block-diagonal components of the
matrix $\hat g_{ch}$ are now
\begin{equation}\label{}
\hat g_{\psi \psi } 
= \left[\begin{array}{cc}
-i \pi\rho & - 2i\pi \rho (1 - 2 f^{(0)}_{\psi})\cr 0 &  i \pi \rho 
\end{array} \right]
\end{equation}
and
\begin{equation}\label{}
\hat g_{\varphi \varphi } 
= \left[\begin{array}{cc}
-i \pi\rho & - 2i\pi \rho (1 - 2 f^{(0)}_{\varphi})\cr 0 &  i \pi \rho 
\end{array} \right],
\end{equation}
where
\[
f^{(0)}_{\psi }(\omega)= \alpha^{2}f_{L}(\omega)+ \beta^{2}f_{R}(\omega)
\]
and 
\[
f^{(0)}_{\varphi }(\omega)= \beta^{2}f_{L}(\omega)+ \alpha^{2}f_{R}(\omega)
\]
define weighted-averages of the distribution functions in the two
leads.  The subscripts (0) are included, to delineate these functions
from the renormalized local conduction electron distribution functions that
can, in principle develop when the leads are coupled. 

There are no retarded or
advanced components to the off-diagonal block matrices $\hat g_{ch}$,
but the inter-channel Keldysh Green's function does become finite when
there is a voltage between the two leads, 
\[
\hat g_{\varphi \psi } (\omega) = \hat g_{\psi \varphi} (\omega)
= \left[\begin{array}{cc}
0 & g^{K}_{\varphi\psi}\cr 0 & 0 
\end{array} \right]
\]
where 
\begin{eqnarray}\label{l}
g^{K}_{\psi \varphi }(\omega) = 
g^{K}_{\varphi \psi }(\omega) 
&=& (4  i \pi\rho) \alpha\beta (f_{R}(\omega)-f_{L}(\omega)),
\end{eqnarray}
and in this way, a finite voltage mixes the two channels. 

We shall return to these general expressions later. However, for 
the one-channel problem, only the $\psi $ electrons couple to the
quantum dot, and so our interest now focuses on $g_{\psi \psi }$.

The Dyson equations for the propagators of conduction $\psi$-electrons and slave $\chi$-fermions (Fig.2) are
\begin{eqnarray}
\hat G^{-1}&=&\hat g_{\psi\psi}^{-1}-\hat \Sigma, \label{psidyson}\\
\hat J^{-1}&=&\hat J_0^{-1}-\hat \Pi, \label{jdyson}
\end{eqnarray}
where 
\begin{equation}\label{}
\hat g_{\psi \psi }^{-1}
= \left[\begin{array}{cc}
\frac{1}{-i \pi\rho} & \frac{2}{- i\pi \rho }(1 - 2 f_{\psi})\cr 0 &  \frac{1}{i \pi \rho }
\end{array} \right]
\end{equation}
is the bare inverse conduction propagator and 
\begin{equation}
\hat J_{0}^{-1}=\left(
\begin{array}{cc}
(\omega-E_{d}-\lambda+i\delta ) 
&   0\\
0 &  (\omega-E_{d}-\lambda-i\delta ) 
\end{array}
\right),
\end{equation}
is the inverse propagator of the $\chi $ fermion.
Notice how the Keldysh component of $g_{\psi \psi}^{-1}= \frac{2}{-i
\pi \rho } (1 - 2 f^{(0)}_{\psi} (\omega))$ is {\sl finite}, whereas the
Keldysh component of $J_{0}^{-1}$ is infinitesimal, and has been
dropped from the above equations. 

The self-energies are generated in our large-N mean-field theory
within the non-crossing approximation (NCA). The diagrammatic Dyson
equations are shown on Fig.1. The boson is our large-N approach is
approximated as a sharp excitation, with an average occupancy $\langle
n_{b\sigma \mu}\rangle \equiv n_{B}= (2S K)/ (NK)$ (see Appendix I). 
In all our calculations, we make use of the 
bare Green's function for the Schwinger bosons 
\begin{equation}
\hat B_{0}=\left(
\begin{array}{cc}
 B_{0}^R &  B_{0}^K \\
0 &  B_{0}^A
\end{array}
\right),
\end{equation}
where
\begin{eqnarray}
B_0^{R,A}&=&\frac{1}{\nu-\lambda\pm i\delta} \nonumber \\
B_0^K&=&(B_0^R-B_0^A)h_{B}, 
\end{eqnarray}
where $h_{B}=1+2n_{B} (\lambda)$.

\begin{figure}[t]
\includegraphics[width=0.45\textwidth]{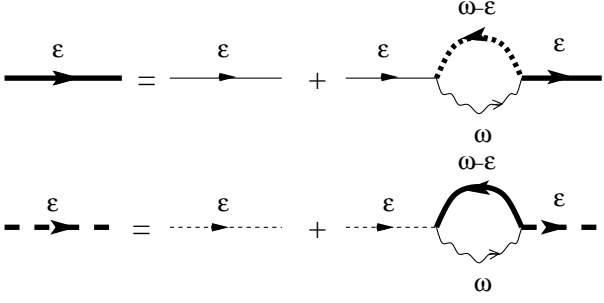}
\caption{The non-crossing approximation for the self-energies of conduction electrons 
and $\chi$ fermions. The solid line denotes the  Larkin Ovchinnikov
matrix propagator for the conduction electrons. The 
dashed line denotes the corresponding Green's function of 
the auxiliary ($\chi $) fermions and the wavy line is the
bosonic propagator. Thin lines denote the bare propagator and full
lines the dressed propagator. Each vertex corresponds to the factor $i\frac{\tilde t}{\sqrt{N}}$.}
\label{NCA}
\end{figure}

>From the Dyson equations we obtain sets of self-consistent integral
equations for the retarded, advanced and Keldysh self-energies (Appendix I). 
The self-energies for the slave ($\chi $) fermion are 
\begin{eqnarray}
\Pi^R(\epsilon)&=&-t^2 n_B(\lambda)G^A(\lambda-\epsilon)\nonumber \\ &-&t^2\int_{-\infty}^{\infty}\frac{d\omega}{\pi}f_{\psi}(\omega)\frac{1}{\omega-\lambda+\epsilon+i\delta}ImG^R(\omega) \nonumber \\
\Pi^K(\epsilon)&=&-2it^2 ImG^R(\lambda-\epsilon)\left[n_B(\lambda)-f_{\psi}(\lambda-\epsilon)\right. \nonumber \\ &-& \left.2n_B(\lambda)f_{\psi}(\lambda-\epsilon) \right].
\label{self_pi}
\end{eqnarray}
The advanced self-energy is determined from the complex conjugate of
the retarded self-energy $\Pi^{A} (\omega)= \Pi^{R} (\omega)^{*}$.
In obtaining these results, we have assumed that the conduction
electron Keldysh Green's function is determined by the relation
\[
G^{K} (\omega) = (G^{R} (\omega) -G^{A} (\omega))h_{\psi}(\omega)
\]
where {\sl \` a priori}, $h_{\psi } (\omega)= 1 - 2 f_{\psi } (\omega)$ 
no longer equal to its equilibrium distribution.

The ratio of the Keldysh to the retarded self-energies
self-consistently determines the $\chi $ fermion distribution functions
\begin{equation}
h_{\chi}\equiv 1 - 2 f_{\chi }=\frac{\Pi^K}{\Pi^R-\Pi^A}=\frac{h_{\psi}h_B-1}{h_B-h_{\psi}}.
\end{equation}
Rearranging this expression gives
\begin{equation}\label{distfn3x}
h_{\psi } (\omega) =
\frac{h_{\chi}h_B-1}{h_B-h_{\chi}}.
\end{equation}

The self-energies for the conduction electrons are 
\begin{eqnarray}
\Sigma^R(\epsilon)&=&-t^2\frac{K}{N} n_B(\lambda)J^A(\lambda-\epsilon)\nonumber \\ &-&t^2\frac{K}{N}\int_{-\infty}^{\infty}\frac{d\omega}{\pi}f_{\chi}(\omega)\frac{1}{\omega-\lambda+\epsilon+i\delta}ImJ^R(\omega) \nonumber \\
\Sigma^K(\epsilon)&=&-2it^2\frac{K}{N} ImJ^R(\lambda-\epsilon)\left[n_B(\lambda)-f_{\chi}(\lambda-\epsilon) \right. \nonumber \\ &-& \left. 2n_B(\lambda)f_{\chi}(\lambda-\epsilon) \right].
\label{self_sigma}
\end{eqnarray}  
and again, $\Sigma ^{A} (\omega)= \Sigma ^{R}
(\omega)^{*}$. The ratio of Keldysh to the retarded/advanced Green's
function is given by
\begin{equation}\label{distfn2x}
h_{\psi } (\omega) =\frac{\Sigma^K}{\Sigma_{\chi }^R-\Sigma_{\chi}^A}=\frac{h_{\chi}h_B-1}{h_B-h_{\chi}}.
\end{equation}
but this recovers exactly the result obtained in \eqref{distfn3x},
showing that detailed balance is satisfied.   However, we
can't choose any distribution function $h_{\psi }$. If we go back to the
original Dyson equation for the conduction electron Green's function
(\ref{psidyson}), the Keldysh component of the Green's
function is given by
\[
G^{K} = G^{R} [\Sigma^{K}- (g_{\psi \psi}^{-1})^{K}]G^{A}
\]
But the distribution function associated with the self-energy is
determined by $\Sigma^{K}= h_{\psi}(\Sigma^R-\Sigma^{A})$ whereas the
distribution function associated with $[g_{\psi\psi}^{-1}]^{K}=  2/ (-i\pi\rho ) 
(1 - 2 f^{(0)}_{\psi })$ is the bare distribution function $h_{\psi
}^{(0)}= 1 - 2 f^{(0)}_{\psi }$. In this way, we see that our original
assumption $G^{K}= (G^{R}-G^{A})h_{\psi }$ requires that $h_{\psi }= h_{\psi}^{(0)}$.
In other words, in large $N$ limit of the single-channel quantum dot 
electron distribution function is unaffected by the coupling to the
dot. 
%Here $J^R(\epsilon)=[\epsilon-E_d-\Pi^R(\epsilon)]^{-1}$ and
%$G^R(\epsilon)=[(-\pi \rho)^{-1}-\Sigma^R(\epsilon)]^{-1}$ are the
%retarded propagators for the $\chi$ fermions and conduction electrons.

We can summarize the results of our calculation of the Keldysh self-energies
by providing the distribution functions that they generate. 
The distribution function of the auxiliary fermion is simply
determined by the relationship 
\begin{equation}
f_{\chi} (\omega) =\frac{n_b[1-f_{\psi} (\omega)]}{n_b+f_{\psi} (\omega)},
\end{equation}
where $n_{b}= 1/ (e^{\beta \lambda}+1)$ determines $\lambda$. 
This relationship can be simply understood as the result of detailed
balance between rate of the decay processes $c\rightarrow b+\chi $ and 
$b+\chi \rightarrow c$, and it reverts to the equilibrium Fermi Dirac
distribution in the limit $V\rightarrow 0$.

\begin{figure}[t]
\includegraphics[width=0.45\textwidth]{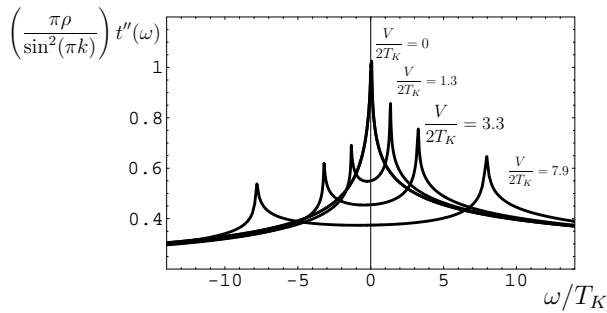}
\caption{Imaginary part of the t-matrix for a variety of voltages for $K/N=0.4$. As the voltage is increased, the singular central peak splits into two components.}
\label{t_matrix_V}
\end{figure}

We have solved the equations
\eqref{self_pi} and \eqref{self_sigma} numerically to obtain 
the energy-dependent $t$-matrix, given by 
\begin{equation}\label{}
\hat t_{\psi } (\omega) =\hat \Sigma(\omega) (1-\hat g_{\psi\psi} (\omega)\Sigma (\omega))^{-1}.
\end{equation}
Actually, the Keldysh part of this quantity is determined simply from
the relation $t_{\psi}^{K}= ( t_{\psi}^{R}-t_{\psi}^{A})h_{\psi
}^{(0)}$, and the advanced part of the electron t-matrix can be simply
written
\begin{equation}\label{advancedt}
t_{\psi }^{A} (\omega) =\frac{\Sigma^{A}(\omega)}{(1-i \pi\rho \Sigma^{A} (\omega))}
\end{equation}
The imaginary part of $t^{A} (\omega)$
determines the electronic density of
states of the resonantly screened d-electrons. The voltage dependence
of this quantity (at  $T=0$) is shown in
Fig.3.  At zero voltage, the t-matrix contains a logarithmic
divergence, which splits into two peaks at a finite voltage.  In this
large $N$ limit, the split Kondo resonance retains its singular
structure.  This is an artifact of taking a large $N$ limit
in which the Schwinger bosons are preserve their sharp spectral structure.

\subsection{Conductance}

The conductance $G=\partial I/\partial V$ is defined by the temperature and
voltage dependent current \cite{MW92}
\begin{equation}\label{}
I (V,T)= \frac{e^{2}}{2\hbar} N\rho \int \frac{d\omega}{\pi} \left[f_L
(\omega)-f_R(\omega) \right]{\rm  Im} t^{A} (\omega)
\end{equation}
where the advanced t-matrix is given by (\ref{advancedt}). 

\begin{figure}[t]
\includegraphics[width=0.45\textwidth]{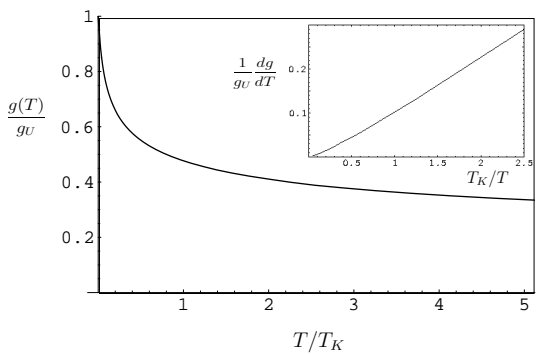}
\caption{Temperature dependence of the differential conductance,
normalized with respect to 
$g_U=N\frac{e^2}{h}sin^2(\pi K/N)$
 for the representative case $K/N=0.4$. Insert shows the $1/T$ divergence of the derivative $dg/dT$.}
\label{cond_temp}
\end{figure}
\begin{figure}[t]
\includegraphics[width=0.45\textwidth]{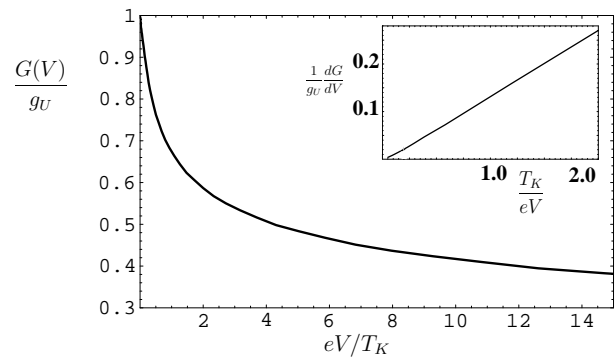}
\caption{Voltage-dependent conductance $G(V)=I(V)/V$ for the case $K/N=0.4$. Insert: derivative of $G(V)$showing $1/V$ divergence.}
\label{cond_volt}
\end{figure}

In Fig.4, we show the computed temperature
dependent conductance. The temperature-dependent deviations from
unitary  conductance are determined by the logarithmic singularity
in the phase shift, and in our calculation, these are proportional to
$1/\ln  (T_{K }/{\rm max} (eV,T))$. There is a technical point here that
needs some discussion. In the Schwinger boson approach, the number of
bound bosons in the Kondo singlet does not exceed $N/2$, which
has the largest bound-state energy, so the region 
$K\geq N/2$,  does not propertly describe an underscreened Kondo
model. Consequently, we are limited to static phase shifts 
 $\delta = \pi (K/N)< \pi/2 $, so the particle-hole symmetric case $\delta =\pi/2$ lies
beyond the reach of this approach. This will leads to some important 
differences between the results of the current calculation and those
expected in experiment. In general our solution does indeed capture
the singular low energy behavior of the phase shift
\begin{equation}\label{}
\delta_{c}\sim \pi k - \frac{\hbox{cons}}{
\ln
\left[\frac{T_{K1}}{T} \right]
}
\end{equation}
Now the conductance $G$ depends on $\sin^{2}\delta (V,T) $, so that
for $k \neq  1/2$
\begin{equation}\label{our}
G\sim \sin^2\delta=\sin^{ 2} (\pi k ) - \frac{\hbox{cons}}{\ln
\left[\frac{T_{K1}}{T} \right]
}
\end{equation}
which is the form that our solutions follow. However, in the 
special case $k=1/2$, the coefficient of leading log in conductance vanishes, 
so the conductance involves square of the logarithm
\[
G\sim \sin^{ 2} (\pi/2 ) - \frac{\hbox{cons}}{\ln^{2}
\left[\frac{T_{K1}}{T} \right]
}
\]
In both cases, the temperature derivative of the
conductance has the singular divergence 
$dG/dT \sim 1/T$, a feature which can be tested experimentally. 

Finally, Fig.5 shows the voltage dependence of the conductance,
which has a similar logarithmic singularity at low voltage. The insert
shows $d^{2}I/dV^{2}$ versus $1/eV$.

\section{Two-channel spin-1 quantum dot: two-stage Kondo effect } \label{sec_twoch}

We start with a derivation of the general current expression for the
case of a two-channel quantum dot.  In this section we consider the
system, described by the Hamiltonian \eqref{And_2ch} and finite
voltage.  We show that at finite voltage the current has three
contributions: contribution from each channel and the interference
term.

\begin{widetext}
\subsection{Derivation of the two-channel current expression}
The current from the dot into the left, and right hand leads is
defined from 
\begin{equation}
I_{L,R}=-e\langle \dot N_{L,R}
\rangle=-\frac{ie}{\hbar}\langle[H,N_{L,R}] \rangle .\label{curr_def}
\end{equation}
The non-zero contribution to the current derives from the commutator
with the hybridization $H_{C}$ in equation (\ref{hybrid}).
In equilibrium $I_{R}=-I_{L}= I$.
The current into the right-hand lead is given by
\begin{eqnarray}\label{expandit}
I_{R }&=& -\frac{i e}{\hbar }\sum_{k\sigma }
\left[\beta t_{1} \langle X_{1\sigma }\dg c_{kR\sigma }\rangle +
\alpha t_{2}\langle 
X_{2\sigma }\dg c_{kR\sigma }\rangle -{\rm  H.c.}
\right]\cr
&=&-\frac{i e}{\hbar }\sum_{k\sigma }
t_{1}\left[ \alpha \beta 
\bigl( \langle X\dg_{1\sigma }\varphi_{k\sigma }\rangle -{\rm H.c.}\bigr)
+\beta^{2}
\bigl( \langle X\dg_{1\sigma }\psi_{k\sigma }\rangle -{\rm H.c.}\bigr)
\right]
\cr
&-& \frac{i e}{\hbar }\sum_{k\sigma }
t_{2}\left[ \alpha \beta 
\bigl( \langle X\dg_{2\sigma }\psi_{k\sigma }\rangle -{\rm H.c.}\bigr)
+\alpha^{2}
\bigl( \langle X\dg_{2\sigma }\varphi_{k\sigma }\rangle -{\rm H.c.}\bigr)
\right].
\end{eqnarray}
Now the expectation of the occupancy of  the dot in each channel is
a constant in the steady state, so that 
\[
\frac{d}{dt}\langle n_{d\lambda }\rangle =\frac{i}{\hbar}\langle
[H_{C},
n_{d\lambda }]\rangle =0, \qquad (\lambda = 1,2)
\]
Carrying out the commutator, we obtain
\begin{equation}\label{}
0 = \frac{i}{\hbar}\sum_{k\sigma }
\left[\langle X\dg _{1\sigma }\psi _{k\sigma }
\rangle  - {\rm H. c} \right]=-\frac{i}{\hbar}\sum_{k\sigma
}\left[\langle X\dg _{2\sigma }
\varphi _{k\sigma }
\rangle  - {\rm H. c} \right]
\end{equation}
From this consideration, we see that the second terms in equation
(\ref{expandit}) identically vanish, so that the current
\begin{equation}\label{}
I = I_{R}= \frac{ie}{\hbar} \alpha\beta \sum_{k\sigma } 
\left[  t_1 (\langle X_{1\sigma }\dg  \varphi_{k\sigma } \rangle  +
{\rm H. c.})
+ t_2 (\langle 
 X_{2\sigma }\dg  \psi_{k\sigma }  \rangle  + {\rm H. c.}]
\right]
\end{equation}
A similar procedure for $J_{L}$ confirms that $J_{L}=-I$.
Note, that in contrast to the Hamiltonian, 
the current is determined by the overlaps between 
the $\psi$-field with the $d_2$-electron and the $\varphi$-field with
the $d_1$-electron on the dot.

\end{widetext}

The expectation values that enter into current are directly related to
the equal-time Keldysh Green functions, 
\begin{eqnarray}\label{l}
\sum_{k\sigma }\langle \varphi_{k\sigma }\dg (t) X_{1\sigma } (t)\rangle
=\frac{N}{2i}
\int \frac{d\omega}{2 \pi } G_{1\varphi}^K (\omega)
\end{eqnarray}
and
\begin{eqnarray}
\sum_{k\sigma }\langle \psi_{k\sigma }\dg (t) X_{2\sigma } (t)\rangle
=\frac{N}{2i}
\int \frac{d\omega}{2 \pi } G_{2\psi}^K (\omega),
\end{eqnarray}
enabling us to write the current in the form
\begin{eqnarray}
I=\frac{Ne}{2\hbar}\alpha \beta\int \frac{d\omega}{2\pi} \left[ t_1(G^K_{\varphi 1}(\omega)-G^K_{1\varphi}(\omega)) \right. \nonumber \\ \left.+t_2(G^K_{\psi 2}(\omega)-G^K_{2\psi}(\omega))\right].
\label{curr_2ch}
\end{eqnarray}

Combining the Dyson equations for the Green's functions $G_{1\varphi}$ and
$G_{2\psi}$ with the expressions for bare Green's functions from
Section IVB (for details see Appendix II) we 
obtain the general current expression in case of two channels 
\begin{equation}
I=N ( I_{\varphi}+I_{\psi}+I_{int}),
\end{equation}
where
\begin{eqnarray}
I_{\psi}=\frac{2e}{\hbar} (\alpha\beta)^2 t_1^2 \rho \int d\omega (f_L(\omega)-f_R(\omega))ImD_{11}^A \nonumber \\
I_{\varphi}=\frac{2e}{\hbar} (\alpha\beta)^2 t_2^2 \rho \int d\omega (f_L(\omega)-f_R(\omega))ImD_{22}^A,
\label{curr_sa}
\end{eqnarray}
where $D_{ij}(t-t')=\langle T_{cK} X_{i\sigma }(t)X_{j\sigma }\dg (t') \rangle$ is the Green's function of the electrons on the dot.

The interference between the channels is induced by the terms 
\begin{eqnarray}
I_{int}=-\frac{e}{2\hbar} \alpha \beta  t_1t_2 i \rho \int d\omega \{ D_{21}^K+D_{12}^K \nonumber \\ +(1-2\beta^2f_L(\omega)-2\alpha^2f_R(\omega))(D_{21}^A-D_{12}^R) \nonumber \\ +(1-2\alpha^2 f_L(\omega)-2\beta^2 f_R(\omega))(D_{12}^A-D_{21}^R) \}
\label{inter}
\end{eqnarray}
Notice
that in the one-channel limit,  ($t_2=0$), the total current
reduces to 
\begin{equation}\label{}
I = NI_\psi =\frac{Ne}{\hbar} \int d\omega [f_L(\omega)-f_R(\omega)]
\frac{\Gamma}{\pi}ImD_{11}^A (\omega),
\end{equation}
recovering the Meir-Wingreen \cite{MW92} 
expression for the current in a  one channel quantum dot, 
with $\Gamma=\Gamma_L \Gamma_R /(\Gamma_L+\Gamma_R)$, where $\Gamma_L=2\pi \rho\alpha^2 t_1^2$ and $\Gamma_R=2\pi\rho \beta^2 t_1^2$.

\subsection{Approximate expression for the  conductance}

In order to calculate the conductance, we need the Dyson equations for
the Green's functions, defined by  Hamiltonian  (\ref{main}).
The Green's functions are defined in the same way as for the one-channel case and the Dyson equations (Fig 6) are
\begin{equation}
\left(
\begin{array}{cc}
\hat G_{\psi\psi} & \hat G_{\psi\varphi} \\
\hat G_{\varphi\psi} & \hat G_{\varphi\varphi}
\end{array}
\right)^{-1}=
\left(
\begin{array}{cc}
\hat g_{\psi\psi} & \hat g_{\psi\varphi} \\
\hat g_{\varphi\psi} & \hat g_{\varphi\varphi}
\end{array}
\right)^{-1}-
\left(
\begin{array}{cc}
\hat \Sigma_{\psi\psi} & \hat \Sigma_{\psi\varphi} \\
\hat \Sigma_{\varphi\psi} & \hat \Sigma_{\varphi\varphi}
\end{array}
\right)
\label{dyson_g}
\end{equation}
\begin{equation}
\left(
\begin{array}{cc}
\hat J_{11} &\hat J_{12} \\
\hat J_{21} & \hat J_{22}
\end{array}
\right)^{-1}=
\left(
\begin{array}{cc}
(\hat J_{0}^{-1})_{11}
& 0 \\
0 & \hat (\hat J_{0}^{-1})_{11}
\end{array}
\right)-
\left(
\begin{array}{cc}
\hat \Pi_{11} & \hat \Pi_{12} \\
\hat \Pi_{21} & \hat \Pi_{22}
\end{array}
\right)
\label{dyson_j}
\end{equation}
Here
\begin{eqnarray}
(J_{0}^{-1})_{ii}^{R,A}
&=&{\omega-E_i-\lambda\pm i\delta} \nonumber \\
(J_{0}^{-1})_{ii}^{K}&=&0.
\end{eqnarray}
Unlike the single channel case, the 
solution of these equations at finite voltage is 
complicated by the need to self-consistently compute the 
distribution functions.  This is because the hybridization no longer
commutes with the Keldysh Green's functions of the conduction sea.
In the following we limit our calculation to the linear response at
zero voltage, which can be written in terms of the equilibrium  $V=0$
Green's functions. 

\begin{figure}[!ht]
\begin{center}
\includegraphics[width=0.5\textwidth]{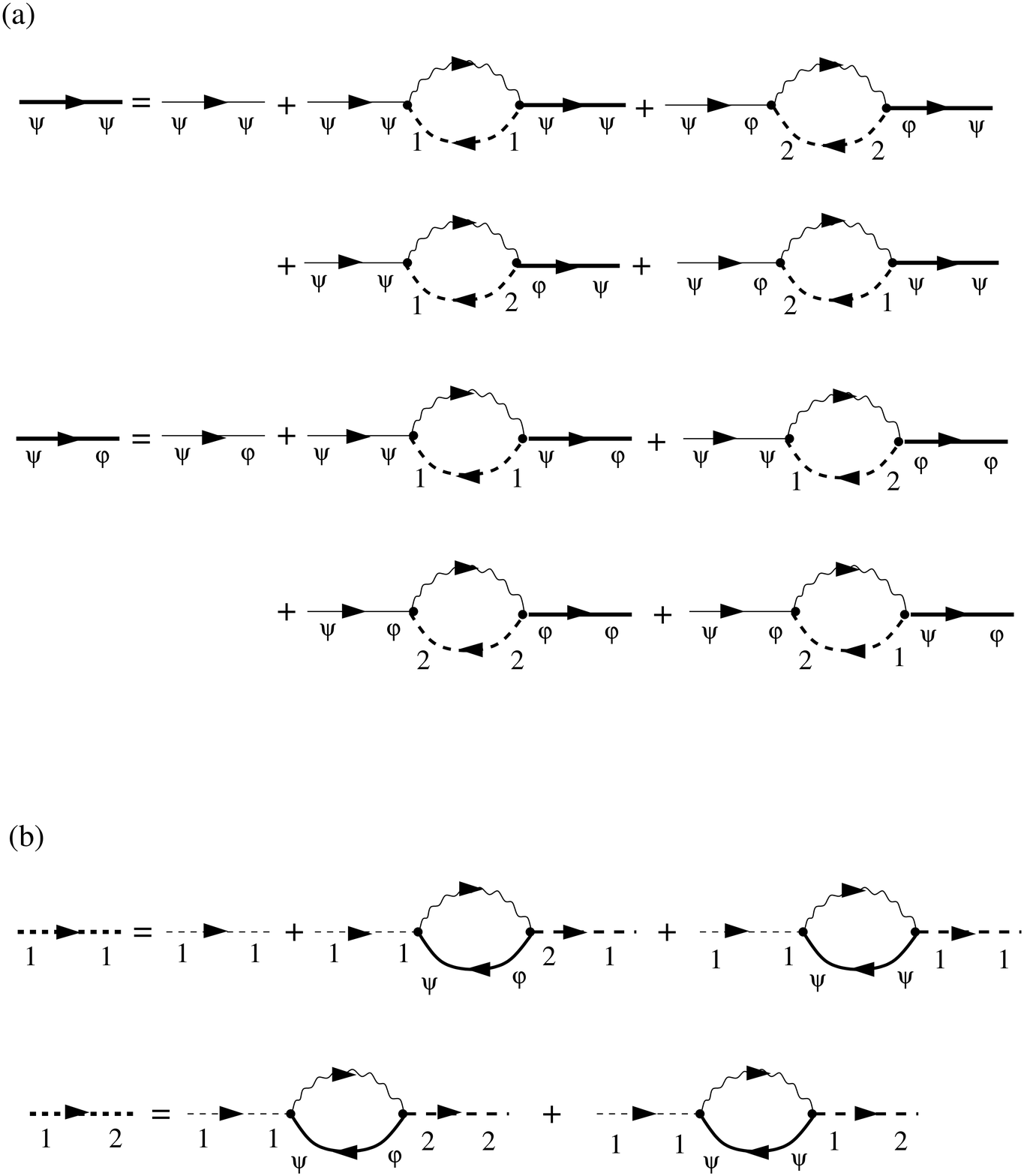}
\end{center}
\caption{Dyson equations for the conduction electron propagators (a)
and the slave fermion propagators (b) for the two-channel quantum dot.}
\label{dyson_eq}
\end{figure} 

In equilibrium at $V=0$, the  Green's functions are channel-diagonal
\begin{eqnarray}
\hat G_{\psi\psi}^{-1}&=&\hat g_{\psi\psi}^{-1}- \hat \Sigma_{\psi\psi} , \nonumber \\
\hat G_{\varphi\varphi}^{-1}&=& \hat g_{\varphi\varphi}^{-1}- \hat \Sigma_{\varphi\varphi}, \nonumber \\
\hat J_{11}^{-1}&=&\hat (J_{11}^0)^{-1}-\hat \Pi_{11}, \nonumber \\
\hat J_{22}^{-1}&=&\hat (J_{22}^0)^{-1}-\hat \Pi_{22}.
\label{diag_terms}
\end{eqnarray}
In this case the single-channel contributions to the current \eqref{curr_sa} can be expressed via ``channel'' t-matrices
\begin{eqnarray}\label{channelI}
I_{\psi}=\frac{2e\rho }{\hbar} (\alpha\beta)^2 \int d\omega (f_L(\omega)-f_R(\omega))ImT_{\psi} (\omega) \\
I_{\varphi}=\frac{2e\rho }{\hbar} (\alpha\beta)^2 \int d\omega (f_L(\omega)-f_R(\omega))ImT_{\varphi} (\omega),
\label{curr_single_linear}
\end{eqnarray}
where
$T_{\psi } =\Sigma_{\psi\psi}^A(1-i\pi\rho\Sigma_{\psi\psi}^A)^{-1}$,
and 
$T_{\varphi}=\Sigma_{\varphi\varphi}^A(1-i\pi\rho\Sigma_{\varphi\varphi}^A)^{-1}$
are the advanced t-matrices.
In linear response the single-channel t-matrices are proportional to the diagonal Green's function of the electrons on the dot holds a simple relation between channel Green's functions and the Green's functions of the physical electrons on the dot
\begin{eqnarray}
T_{\psi} (\omega)&=& ( t_{1})^{2} D^{A}_{11} (\omega), \nonumber \\
T_{\varphi} (\omega)&=&( t_{2})^{2}D^{A}_{22} (\omega).
\end{eqnarray} 
where
\begin{equation}
D^{A}_{\lambda \lambda'} (\omega) = i \int_{-\infty }^{0}\langle 
\{X_{\lambda \sigma } (t),\ X\dg _{\lambda \sigma } (0)\}\rangle e^{i
\omega t}dt
\end{equation}
and $X\dg _{\lambda\sigma }= \frac{1}{\sqrt{2S}}\sum_{\mu=1,K}b\dg_{\sigma \mu}\chi _{\mu}$.
In the large $N$ limit, the t-matrices for each channel completely
decouple in equilibrium, and are given by the scaling form
\begin{equation}\label{scalingf}
T_{\lambda} (\omega,T)= {\cal T} (\frac{\omega}{T_{K\lambda}}, \frac{T}{T_{K\lambda}})
\end{equation}
where $T_{K\lambda}$ ($\lambda=1,2$) are the Kondo temperatures in the
two channels. 

The interference contributions to the current pose a greater challenge
and we have been unable to treat these terms without  making some
additional degree of approximation. 
Without loss of generality, the Keldysh components of the dot Green's
functions can be written
\[
D^{K} =  D^{R} (-D^{-1})^{K}D^{A}
\]
In linear response, the effect of the voltage on will appear through
the voltage dependence of $( D^{-1})^{K}$, so we can replace the
retarded and advanced combinations in the above by their channel
diagonal components, 
\begin{eqnarray}
D_{12}^K&=&D_{11}^R 
(-D^{-1})^{K}_{12} 
D_{22}^A \nonumber \\
D_{21}^K&=&D_{22}^R
(-D^{-1})^{K}_{21} 
D_{11}^A,
\end{eqnarray}
In general, the Keldysh components of $D^{-1}_{12}$ and
$D^{-1}_{21}$ contain a hybridization component and a term derived
from interactions
\begin{eqnarray}\label{l}
( -D^{-1}_{12})^{K} &=& t_{1}g_{\psi \varphi}^{K }t_{2} - (t_{1}t_{2})\Sigma^{K}_{12}\cr
( -D^{-1}_{21})^{K} &=& t_{2}g_{\varphi\psi }^{K }t_{1} - (t_{1}t_{2})\Sigma^{K}_{21}
\end{eqnarray}
The second terms are a kind of vertex correction to the current
operator (see Fig. 10 (a)).   These terms describe the inelastic
corrections to the interference contribution to the current. 
Our approximation entails neglecting 
these vertex corrections, so that 
(see Appendix III)
\begin{eqnarray}
D_{12}^K&=&t_1t_2D_{11}^Rg_{\psi\varphi}^KD_{22}^A \nonumber \\
D_{21}^K&=&t_1t_2D_{22}^Rg_{\varphi\psi}^KD_{11}^A,
\end{eqnarray}
When the same vertex corrections are ignored, the retarded and advanced Green's
functions are channel-diagonal, $D_{12}^{R,A}=D_{21}^{R,A}=0$.  

With these approximations, 
the contribution to the current due to the interference between the two channels becomes
\begin{eqnarray}
I_{int}=-\frac{2e}{\hbar}(\alpha\beta t_1 t_2)^2\pi \rho^2\int d\omega(f_L(\omega)-f_R(\omega)) \nonumber \\
\left[ D_{11}^R (\omega) D_{22}^A (\omega) +D_{11}^A (\omega)  D_{22}^R (\omega) \right],
\end{eqnarray}
or,
\begin{eqnarray}\label{interferenceterms}
I_{int}=-\frac{2e}{\hbar}(\alpha\beta )^2\pi \rho^2\int d\omega(f_L(\omega)-f_R(\omega)) \nonumber \\
\left[ T_{1}^{*} (\omega) T_{2} (\omega) +T_{1}(\omega)  T_{2}^* (\omega) \right].
\end{eqnarray}
This expression brings out the interference character of the term, and
combined with (\ref{channelI}) describes the linear
response current flow, 
purely in terms of the channel diagonal, equilibrium t-matrics.

We can further simplify the full expression for the linear response
current by taking advantage of the optical theorem. 
In the case where the scattering off the impurity is purely elastic,
the channel diagonal t-matrices satisfy an optical theorem
${\rm Im}T_{\lambda} (\omega)= \pi \rho \vert
T_{\lambda} (\omega)\vert^{2}$.  The deviation from the optical
theorem describes a decoherence scattering rate\cite{decoherence} as follows
\begin{equation}\label{}
\tau^{-1}_{\lambda}= {\rm Im}T_{\lambda} (\omega)- \pi \rho \vert
T_{\lambda} (\omega)\vert^{2}.  
\end{equation}
In this form, the combined interference and channel diagonal currents
can be written in the simple form
\[
G = G_{coh}+ G_{inc}
\]
where
\begin{equation}\label{gcoherent}
G_{coh}=\frac{Ne^{2}}{h}(2\alpha\beta )^2\int d\omega
\left(- \frac{df (\omega)}{d\omega} \right)
\nonumber \\
\vert \pi\rho ( T_{1} (\omega)- T_{2} (\omega))\vert^{2}
\end{equation}
defines the coherent conductance through the two-channel and
\begin{eqnarray}\label{gincoherent}
G_{inc}&=&\frac{Ne^{2}}{h}(2\alpha\beta )^2 
\int d\omega
\left(- \frac{df (\omega)}{d\omega} \right)\cr
&\times& \pi\rho[\tau^{-1}_{1} (\omega)+\tau^{-1}_{2} (\omega)]
\end{eqnarray}
the additional contribution due to incoherent transport.
In the limits where the scattering t-matrix is purely elastic,
\[
T_{\lambda} (\omega) = \frac{e^{i\delta_{\lambda} (\omega)}\sin
\delta_{\lambda} (\omega)}{\pi\rho }
\]
so that the inelastic scattering rate $\tau_{\lambda}^{-1}=0$ so at 
low temperatures
\begin{eqnarray}\label{l}
G_{coh} &=& \frac{ Ne^{2}}{h} (2 \alpha \beta)^{2} (\pi\rho)^{2}\vert T_{1} (0) -T_{2} (0)\vert^{2}
\cr
&=&\frac{ Ne^{2}}{h} (2 \alpha \beta)^{2}\sin^2(\delta_{1}-\delta_{2})
\end{eqnarray}
recovering the  result expected from Landauer theory.

In the large $N$ limit it is straightforward to compute this
approximate expression for the conductance. The scaling form (\ref{scalingf})
can be used to rescale the t-matrix from one channel to the
other. Thus if $T_{K1}/T_{K2}=\alpha $, then from (\ref{scalingf}), 
\[
T_{2} (\omega,T)= T_{1} (\alpha \omega, \alpha T).
\]
The behavior of the linear conductance in case of the two-channel
quantum dot for different ratios of the two Kondo temperatures is
shown on Fig. 7. When the two Kondo temperatures are equal to each
other, the conductance is completely suppressed due to destructive
inter-channel interference. When $T_{K1}$ and $T_{K2}$ are widely
separated, at low temperatures conductance is suppressed as expected,
and at high temperatures it develops a characteristic hump-structure
due to the Kondo resonance in one of the channels.
\begin{figure}[t]
\includegraphics[width=0.45\textwidth]{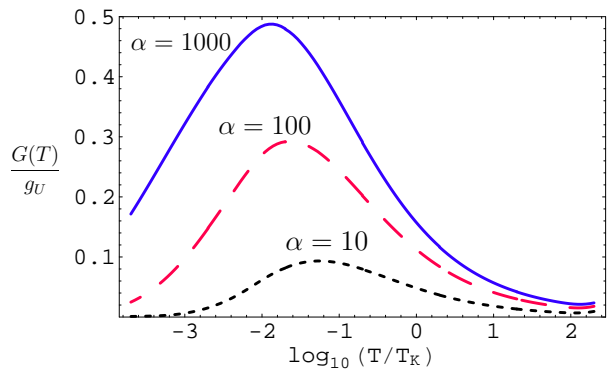}
\caption{Linear conductance of the two-channel quantum dot normalized with
respect to $g_U=N\frac{e^2}{h}sin^2(\pi K/N)$, 
calculated for $K/N=0.4 $ and  $\alpha=\beta=1/\sqrt{2}$, 
using
equations (\ref{gcoherent}) and (\ref{gincoherent}), 
for three ratios of the two Kondo temperatures:
$\alpha =T_{K1}/T_{K2}=10$
(short-dashed curve), $\alpha =T_{K1}/T_{K2}=100$
(long-dashed curve) and $\alpha =T_{K1}/T_{K2}=1000$
(solid curve).}
\label{cond_2ch}
\end{figure}

\section{Conclusion and discussion.}

This article has considered the physics of a  quantum dot in which Hund's
coupling between the electrons localized within the dot gives rise to 
a spin-1 configuration on the quantum dot.  Theoretically, such a
quantum dot  is expected to map onto a two-channel, spin-1 Kondo model
in which a $\pi/2$ phase shift develops in both scattering
channels. According to a Landauer analysis of the resulting Fermi
liquid, the transmission through the dot should vanish at low
temperatures, giving rise to a  conductance
\begin{equation}\label{}
G = 2\frac{e^2}{h} \sin ^2 ( \delta_1 - \delta_2)
\end{equation}
that vanishes at low temperatures
\cite{Pustilnik01,zarand04}. Paradoxically, experiment suggests that
spin-1 quantum dots do indeed develop a zero-bias anomaly\cite{schmid00,sasaki00,kogan03}. 
It is this issue that has motivated the current body of work. 

Motivated by these results, the current authors proposed\cite{Anna05}  that triplet 
quantum dots with a large zero bias correspond to anisotropic spin-1
Kondo models, in which the Kondo temperatures of the two channels
exhibit a large ratio.  In this paper, we have explored the general
problem of high-spin quantum dots, screened by two conduction
channels, attempting to develop a framework to model the detailed
conductance of such systems. 
Our proposed model involves an infinite $U$ description of the electrons in the
quantum dot, with an infinitely strong Hund's interaction that ensures
the formation of a high-spin electron configuration in the dot.
In this case the ground state of the dot is a spin-triplet and the charge fluctuations
to the excited singly occupied states can be eliminated via
Schrieffer-Wolff transformation. The Hamiltonian which describes the
spin-1 quantum dot at low temperatures is then that of a two-channel
Kondo Hamiltonian with two different Kondo coupling constants,and two
different Kondo temperatures for the two electron states inside the triplet. 

The emergence of two scattering channels in a spin-1 quantum dot has 
interesting consequences: in particular, the conductance develops a 
a non-monotonic dependence on magnetic field, temperature and 
voltage \cite{Pustilnik01}. At zero temperature, destructive
interference between the two elastic scattering channels leads to a
complete suppression of conductance and the resulting ground state 
is described within a Landau Fermi liquid picture. 
Conductance starts gradually increasing with temperature and
between $T_{K2}$ and $T_{K1}$ reaches it's maximum (the height of the
maximum which is less or equal to the unitary conductance, depends on
the difference between the two Kondo temperatures). At still higher 
temperatures  $T>>T_{K1}$, the conductance is ultimately
suppressed to zero. Note that when the two Kondo temperatures are
equal to each other, conductance remains suppressed for all
temperatures\cite{zarand04}.

Our analytic analysis of this non-monotonic dependence
of conductance on temperature employs a 
a Schwinger boson formalism to describe the dot spin degrees of
freedom,  using a large-N expansion to provide an approximate
treatment of the resulting many body physics. 
Although the bosons in our
approach remain unrenormalized, which means we neglect the spin-decoherence
effects we find that the that the initial increase,
and ultimate suppression of conductance are reproduced in our
analysis.

As part of this work, we considered the interesting case when one of the scattering
channels is completely suppressed, which provides a realization of
the underscreened Kondo effect in the spin-1 quantum dot. In
this case at zero temperature conductance reaches its unitary limit,
but in a specific singular way. Although the scattering shift at $T=0$
is equal to $\delta=\pi/2$, the energy dependence of $\delta$ contains
logarithmic contributions. This situation is described in terms of
singular Fermi liquid \cite{pankaj}.  It turns out that our Schwinger
boson approximation is suited for description of a singular behavior
of conductance at low temperatures and low voltages.

There are several open questions, of both a theoretical, and
experimental nature that arise from this work. The current through the
one-channel quantum dot is compactly related to the density of states
of the dot-electrons via the Meir and Wingreen relation. In the
two-channel case, these simplifications continue to operate in the
contributions to the  current that are channel diagonal,
but the off-diagonal components to the current require a knowledge of
the Keldysh Green's functions and the voltage-induced components to
interchannel t-matrix. Is the physical reason for this added
complexity? One possibility, that we have not been able to eliminate, 
is that a more general matrix ansatz for the Keldysh
Green's functions, of the form, such as 
\[
G^{K} (\omega) = G^{R} (\omega) F (\omega) - F (\omega)G^{A} (\omega)
\]
can be used to simplify our results. 

On the experimental front, we are still lacking hard evidence that the
zero-bias anomaly in  triplet dots is driven by an extreme ratio of
Kondo temperatures.  It would be particularly interesting if lateral
quantum dots of the kind used in previous experiments could be adapted
to support a variable ratio between the Kondo temperatures in the two
channels. This might be done, for example, through the introduction of
additional gates that change the symmetry of the quantum dot.  This
might provide the means to observe the cross-over from the singular
Fermi liquid associated with fully developed
underscreened behavior, to the interference dominated, non-monotonic
conductance of the fully screened quantum dot.

The authors wish to thank H. Kroha, G. Schon and G. Zarand 
for discussions and comments related to this work. We are particularly indebted to M. Eschrig for many useful discussions.
 This research was partly supported
by DFG CFN (AP, BB)
and DOE grant DE-FG02-00ER45790 (PC).

\section*{Appendix I: Self-energies for the one channel Kondo effect.}

The self-energy contributions to the equations \eqref{psidyson} and \eqref{jdyson} are
shown in Fig.8. Each vertex is associated with a factor
$\frac{i \tilde{t}}{\sqrt{N}}$. Internal summations over the spin index, or
replica indices inside the loop provide a factor of $N$ or $K$
respectively, so that the conduction and slave fermion self energies
are of order $O (N/N)=O (1)$ and $O (K/N)$ respectively, and they
remain finite in the large $N$ limit. By contrast,
the bosonic self-energy contains no internal summation over internal quantum
numbers, so that this quantity is of order $O (1/N)$ and vanishes in
the large $N$ limit.

The conduction electron self energy is directly related to the Green's
function of the d-electron, written in composite form as
\begin{equation}\label{}
\Sigma_{\pm \pm} (t-t') = \frac{-i \tilde{t}^{2}}{N} \sum_{\mu}
\langle T_{c_{K}} \left(\chi \dg b_{\sigma \mu} \right) (t)
\left(
b\dg _{\sigma \mu} \chi \dg \right) (t')\rangle 
\end{equation}
where $T_{c_{K}}$ refers to time-ordering along the Keldysh contour 
and the notation $A_{\pm}\equiv A_{1}\pm A_{2}$ denotes the symmetric
and antisymmetric combination of fields from the forwards time (1) and
backwards time (2) parts of the contour. (Sometimes $A_{+}\equiv
A_{cl}$ and $A_{-}\equiv A_{q}$ are referred to as the classical and
quantum fields, respectively.) 

\begin{figure}[t]
\includegraphics[width=0.4\textwidth]{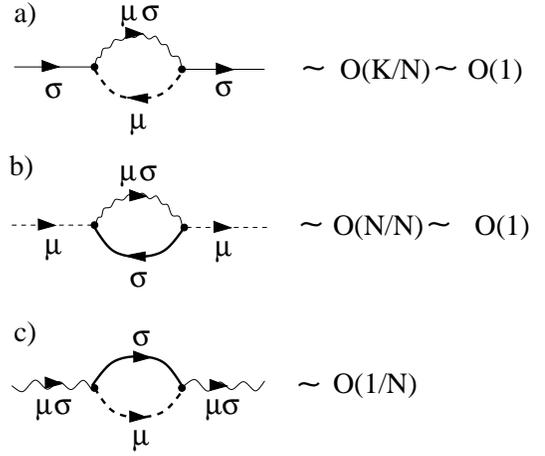}
\caption{NCA contributions to the self energies of a) lead electrons, b) auxiliary $\chi$-holes, and c) Schwinger bosons. The solid line denotes the  Larkin Ovchinnikov
matrix propagator for the conduction electrons. The 
dashed line denotes the corresponding Green's function of 
the auxiliary ($\chi $) fermions and the wavy line is the
bosonic propagator. Each vertex corresponds to the factor $i\frac{\tilde t}{\sqrt{N}}$.}
\label{NCA}
\end{figure}

 The retarded, advanced and
Keldysh combinations of the self energy are related to
\begin{equation}\label{}
\Sigma_{R}\equiv  \Sigma_{+-},\qquad \Sigma_{A}\equiv  \Sigma_{-+},\qquad \Sigma_{K}\equiv  \Sigma_{++}.
\end{equation}
The ``classical'' (+) vertices have an off-diagonal structure in Kedlysh
space, while the quantum  (-) vertices are  diagonal
\[
+ \equiv \frac{1}{\sqrt{2}}\underline{\tau }_{1},\qquad - \equiv \frac{1}{\sqrt{2}}\underline{1}.
\]
With this information, the conduction electron self energy is given by
\[
\Sigma (t)=\mat{\Sigma_R (t)}{\Sigma_{K} (t)}{}{\Sigma_{A} (t)} \equiv \mat{\Sigma_{+-} (t)}{\Sigma_{++} (t)}{0}{\Sigma_{-+} (t)}
\]
which in expanded form gives
\[
\Sigma (t)= i\frac{K}{2N}\mat
{{\rm Tr} \bigl [\tau_{1}B (t)J (-t)\bigr ]}
{{\rm Tr} \bigl [
\tau_{1}B (t)\tau_{1}J (-t)\bigr ]}
{0}
{{\rm Tr} \bigl [
B (t)\tau_{1}J (-t)\bigr ]}
\]
The Keldysh traces give the combinations
\begin{eqnarray}\label{l}
{{\rm Tr} \bigl [\tau_{1}B J \bigr ]}\equiv B_{R}J_{K}+B_{K}J_{A}\cr
{{\rm Tr} \bigl [\tau_{1}B \tau_{1}J \bigr ]}\equiv
B_{R}J_{A}+B_{A}J_{R}+B_{K}J_{K}\cr
{{\rm Tr} \bigl [B \tau_{1}J \bigr ]}\equiv
B_{K}J_{A}+B_{R}J_{K}
\end{eqnarray}
Carrying  out the Fourier transforms, the explicit form for the
conduction self-energy is then
\begin{eqnarray}
\Sigma^R(\epsilon)&=&-it^2\frac{K}{N}\int \frac{d\omega}{4\pi}\left[B^R(\omega)J^K(\omega-\epsilon) \right. \nonumber \\ &+&\left .B^K(\omega)J^A(\omega-\epsilon) \right], \nonumber \\
\Sigma^A(\epsilon)&=&-it^2\frac{K}{N}\int \frac{d\omega}{4\pi}\left[B^K(\omega)J^R(\omega-\epsilon) \right.
\nonumber \\ &+&\left. B^A(\omega)J^K(\omega-\epsilon) \right], \nonumber \\
\Sigma^K(\epsilon)&=&-it^2\frac{K}{N}\int \frac{d\omega}{4\pi}\left[B^K(\omega)J^K(\omega-\epsilon) \right.
\nonumber \\ &+&\left. B^A(\omega)J^R(\omega-\epsilon)+B^R(\omega)J^A(\omega-\epsilon) \right]
\label{onech_sigma}.
\end{eqnarray}
Repeating the same procedure for the slave fermions, we obtain
\begin{eqnarray}
\Pi^R(\epsilon)&=&-it^2\int \frac{d\omega}{4\pi}\left[B^R(\omega)G^K(\omega-\epsilon) \right. \nonumber \\ &+& \left. B^K(\omega)G^A(\omega-\epsilon) \right], \nonumber \\
\Pi^A(\epsilon)&=&-it^2\int \frac{d\omega}{4\pi}\left[B^K(\omega)G^R(\omega-\epsilon) \right. \nonumber \\ &+& \left. B^A(\omega)G^K(\omega-\epsilon) \right], \nonumber \\
\Pi^K(\epsilon)&=&-it^2\int \frac{d\omega}{4\pi}\left[B^K(\omega)G^K(\omega-\epsilon) \right. \nonumber \\ &+& \left. B^A(\omega)G^R(\omega-\epsilon)+B^R(\omega)G^A(\omega-\epsilon) \right].
\label{onech_pi}
\end{eqnarray}
Since the Schwinger boson self-energy vanishes in the large $N$ limit,
the Schwinger boson fields
are unrenormalized and take their bare values
\begin{eqnarray}
B^{R,A}(\nu)&=&\frac{1}{\nu-\lambda\pm i\delta} \nonumber \\
B^K(\nu)&=&-2\pi i \delta(\nu-\lambda)h_{B}
\end{eqnarray}
where $h_{B}=1+2n_{B}$ and $n_B\equiv n_{B} (\lambda)= 2SK / (NK)$ is the Bose distribution function.
The Keldysh Green's function of the fermions can be related to
their retarded and advanced Green functions via the relations
\begin{eqnarray}\label{l}
J^{K} (\omega) &=& (J^{R} (\omega)-J^{A} (\omega)) h_{\chi }
(\omega)\cr
&=& 2 i Im [J_{R} (\omega)] (1- 2 f_{\chi} (\omega)), \cr\cr
G^{K} (\omega) &=& (G^{R} (\omega)-G^{A} (\omega)) h_{\Psi }
(\omega)\cr
&=& 2 i Im [G_{R} (\omega)] (1- 2 f_{\Psi} (\omega)).
\end{eqnarray}
With these simplifications, we can expand the self-energies in the
following form
\begin{eqnarray}
\Pi^A(\epsilon)&=&-t^2 n_B(\lambda)G^R(\lambda-\epsilon) \nonumber \\ &-&t^2\int_{-\infty}^{\infty}\frac{d\omega}{\pi}f_{\Psi}(\omega)\frac{1}{\omega-\lambda+\epsilon-i\delta}ImG^R(\omega) \nonumber \\
\Pi^R(\epsilon)&=&-t^2 n_B(\lambda)G^A(\lambda-\epsilon)\nonumber \\ &-&t^2\int_{-\infty}^{\infty}\frac{d\omega}{\pi}f_{\Psi}(\omega)\frac{1}{\omega-\lambda+\epsilon+i\delta}ImG^R(\omega) \nonumber \\
\Pi^K(\epsilon)&=&-2it^2 ImG^R(\lambda-\epsilon)\left[n_B(\lambda)-f_{\Psi}(\lambda-\epsilon)\right. \nonumber \\ &-& \left.2n_B(\lambda)f_{\Psi}(\lambda-\epsilon) \right]
\end{eqnarray}
The slave fermion distribution function is given by
\begin{equation}\label{distfn}
h_{\chi}=\frac{\Pi_{\chi}^K}{\Pi_{\chi}^R-\Pi_{\chi}^A}=\frac{h_{\Psi}h_B-1}{h_B-h_{\Psi}}.
\end{equation}
Rearranging this expression gives
\begin{equation}\label{distfn3}
h_{\psi } (\omega) =
\frac{h_{\chi}h_B-1}{h_B-h_{\chi}}.
\end{equation}
Similarly, 
\begin{eqnarray}
\Sigma^A(\epsilon)&=&-t^2 \frac{K}{N}n_B(\lambda)J^R(\lambda-\epsilon)\nonumber \\ &-&t^2\frac{K}{N}\int_{-\infty}^{\infty}\frac{d\omega}{\pi}f_{\chi}(\omega)\frac{1}{\omega-\lambda+\epsilon-i\delta}ImJ^R(\omega) \nonumber \\
\Sigma^R(\epsilon)&=&-t^2\frac{K}{N} n_B(\lambda)J^A(\lambda-\epsilon)\nonumber \\ &-&t^2\frac{K}{N}\int_{-\infty}^{\infty}\frac{d\omega}{\pi}f_{\chi}(\omega)\frac{1}{\omega-\lambda+\epsilon+i\delta}ImJ^R(\omega) \nonumber \\
\Sigma^K(\epsilon)&=&-2it^2\frac{K}{N} ImJ^R(\lambda-\epsilon)\left[n_B(\lambda)-f_{\chi}(\lambda-\epsilon) \right. \nonumber \\ &-& \left. 2n_B(\lambda)f_{\chi}(\lambda-\epsilon) \right]
\label{self_2ch}
\end{eqnarray}
We can also compute the electron distribution function from 
\begin{equation}\label{distfn2}
h_{\psi } (\omega) =\frac{\Sigma^K}{\Sigma_{\chi }^R-\Sigma_{\chi}^A}=\frac{h_{\chi}h_B-1}{h_B-h_{\chi}}.
\end{equation}
but this recovers exactly the result obtained in (\ref{distfn3}),
showing that detailed balance is satisfied.

\section*{Appendix II: on the derivation of general current formula in case of two channels}

As an example we present the derivation of the Green's function $G_{1\varphi}$
\begin{equation}
G_{1\varphi}(t_1,t_2)=-i \langle T_{ck} d_1(t_1)\varphi^+(t_2) \rangle,
\end{equation} 
and the Dyson Eq is that depicted on Fig.4 
\begin{figure}[!ht]
\begin{center}
\includegraphics[width=0.5\textwidth]{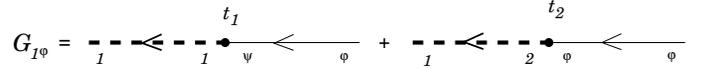}
\end{center}
\caption{The Dyson Eq. for $\hat G_{1\varphi}$ Green's function ($\hat G_{1\varphi}$ is a matrix in Keldysh space).}
\label{tmatrix}
\end{figure} 
To avoid double counting we renormalize only the $d$-electron Green's
function, while the channel Green's function is left unrenormalized, so that the expression for $G_{1a}$ is
\begin{eqnarray}
G_{1\varphi}(t_1,t_2)=\int_{cK} d\tau \left( \hat D_{11}(t_1,\tau)t_1\hat g_{\psi\varphi}(\tau,t_2) \right.\nonumber \\ \left. +\hat D_{12}(t_1,\tau)t_2\hat g_{\varphi\varphi}(\tau,t_2)\right)
\label{g1a_kel}
\end{eqnarray}
where 
\begin{equation}
D_{ij}(t_1,t_2)=-i \langle T_{ck} d_i(t_1)d_j^+(t_2) \rangle.
\end{equation}
Expression \eqref{g1a_kel} can be rewritten in Larkin-Ovchinnikov
space and the Keldysh part of the Green's function can be found. After
Fourier transforming, we obtain the following Keldysh component
\begin{equation}\label{}
G^K_{1\varphi}=t_1(D_{11}^Rg_{\psi\varphi}^K+D_{11}^Kg_{\psi\varphi}^A)+t_2(D_{12}^Rg_{\varphi\varphi}^K+D_{12}^Kg_{\varphi\varphi}^A)
\end{equation}

The Dyson Eq for $G_{2\psi}$ is derived in an analogous fashion
\begin{equation}\label{}
G^K_{2\psi}=t_2(D_{22}^Rg_{\varphi\psi}^K+D_{22}^Kg_{\varphi\psi}^A)+t_1(D_{21}^Rg_{\psi\psi}^K+D_{21}^Kg_{\psi\psi}^A)  
\end{equation}
These equations can now be expanded using the explicit expressions for
the bare channel Green's functions given in  section IV,B to obtain
the general expressions \eqref{curr_sa} and \eqref{inter}. 

\section*{Appendix III: estimation of the interference current term}

\begin{figure}[!ht]
\begin{center}
\includegraphics[width=0.35\textwidth]{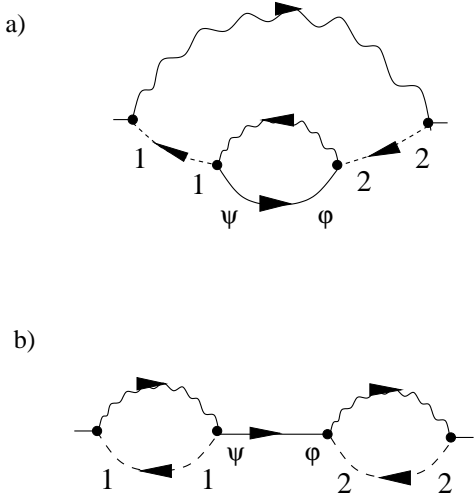}
\end{center}
\caption{Diagrammatic contributions to 
the off-diagonal Green's function $D_{12} (\omega)$, describing
inter-channel contributions to the conduction electron propagator. }
\label{dcontr}
\end{figure} 

Contributions to the current which are produced by channel
interference are defined by the interchannel Green's functions
$D_{12}$ and $D_{21}$. These Green's functions are proportional to the
self-energies of $G$-functions. On Fig. 10 we show two types of
contributions which arise in the first order of the perturbation
theory. Diagrams of type a) contribute to the non-elastic scattering
of d-electrons, and therefore constitute the non-Fermi liquid
corrections to the conductance. Other diagrams can be summed in a
simple Dyson equation 
\begin{eqnarray}
D_{12}=D_{11}^0t_1G_{\psi 2}=D_{11}^0t_1g_{\psi\psi}t_1D_{12}+D_{11}^0t_1g_{\psi \varphi}t_2D_{22}, \nonumber \\
D_{12}=t_1t_2(1-t_1^2D_{11}^0g_{\psi\psi})^{-1}D_{11}^0g_{\psi \varphi}D_{22}
\label{d12}
\end{eqnarray}
where, as usual $D_{11}^0$ are bare Green's functions, $g_{\psi\psi}, g_{\psi\varphi}$- bare conduction electron Green's functions, and $D_{12}, D_{22}$ are the fully-renormalized d-electron Green's functions.

Analogously we can write the Dyson Eq for $D_{11}$ 
\begin{eqnarray}
D_{11}&=&D_{11}^0+t_1^2D_{11}^0g_{\psi\psi}D_{11}+t_1t_2D_{11}^0g_{\psi \varphi}D_{21} \nonumber \\
&=&D_{11}^0+t_1^2D_{11}g_{\psi\psi}D_{11}^0+t_1t_2D_{12}g_{\varphi \psi}D_{11}^0
\label{d11}
\end{eqnarray}
We can write from here
\begin{equation}
D_{11}(1-t_1^2g_{\psi\psi}D_{11}^0)=D_{11}^0(1+t_1t_2D_{12}g_{\varphi \psi}),
\end{equation}
so that
\begin{equation}
D_{11}=(1+t_1t_2D_{12}g_{\varphi \psi})D_{11}^0(1-t_1^2g_{\psi\psi}D_{11}^0)^{-1}
\end{equation}
and
\begin{eqnarray}
(1+t_1t_2D_{12}g_{\varphi \psi})^{-1}D_{11}=D_{11}^0(1-t_1^2g_{\psi\psi}D_{11}^0)^{-1}\nonumber \\
=(1-t_1^2D_{11}^0g_{\psi\psi})^{-1}D_{11}^0
\label{d12help}
\end{eqnarray}
Now we can plug \eqref{d12help} into \eqref{d12} and get
\begin{equation}
D_{12}=(1+t_1t_2D_{12}g_{\varphi \psi})^{-1}t_1t_2D_{11}g_{\psi \varphi}D_{22}
\end{equation}
which leads to the equation
\begin{equation}
D_{12}=t_1t_2D_{11}g_{\psi \varphi}D_{22}-t_1t_2D_{12}g_{\varphi \psi}D_{12}.
\label{d12_fin}
\end{equation}
From \eqref{d12_fin} we now can derive that retarded and advanced off-diagonal terms vanish in this approximation.
Namely, for example
\begin{equation}
D_{12}^R=t_1t_2D_{11}^Rg_{\psi \varphi}^RD_{22}^R-t_1t_2D_{12}^Rg_{\varphi \psi}^RD_{12}^R
\end{equation}
is correct and therefore $D_{12}^R=0$, because $g_{\psi \varphi}^R=g_{\varphi \psi}^R=0$. Keldysh components of $D_{12}$ nevertheless do not vanish
\begin{eqnarray}
D_{12}^K&=&t_1t_2(D_{11}^R g_{\psi \varphi}^R D_{22}^K+D_{11}^R g_{\psi \varphi}^K D_{22}^A+D_{11}^K g_{\psi \varphi}^A D_{22}^A)+ \nonumber \\
&+&t_1t_2(D_{12}^Rg_{\varphi \psi}^RD_{12}^K+D_{12}^Rg_{\varphi \psi}^KD_{12}^A+D_{12}^Kg_{\varphi \psi}^AD_{12}^A)                  
\label{12_kel}
\end{eqnarray}
Most of the terms in \eqref{12_kel} vanish and we are left with
\begin{equation}
D_{12}^K=t_1t_2D_{11}^Rg_{\psi \varphi}^KD_{22}^A.
\end{equation}


\begin{thebibliography} {99}

\bibitem{Hewson} A. C. Hewson, {\it The Kondo Problem to Heavy Fermions}, Cambridge University Press, Cambridge, 1993.

\bibitem{Glazman01} L. Kouwenhoven and L. Glazman, Physics World {\bf 14}, 33 (2001).

\bibitem{sasaki00} S. Sasaki, S. De Franceschi, J. M. Elzerman,
W. G. van der Wiel, M. Eto, S. Tarucha, and L. P. Kouwenhoven,
Nature {\bf 405}, 764 (2000).

\bibitem{kogan03} A. Kogan, G. Granger, M. A. Kastner, and D. Goldhaber-Gordon,
Phys. Rev. B {\bf 67}, 113309 (2003).

\bibitem{Nozieres80} Ph. Nozi\`eres and A. Blandin, J. Physique {\bf 41}, 193 (1980).

\bibitem{Andrei84} N. Andrei and C. Destri, Phys. Rev. Lett. {\bf 52}, 364 (1984).

\bibitem{affleck91} I. Affleck, A. W. W. Ludwig, Phys. Rev. Lett. {\bf 67}, 161 (1991).

\bibitem{Cox87} D. L. Cox, Phys. Rev. Lett. {\bf 59}, 1240 (1987).

\bibitem{Andrei04} N. Andrei and C. J. Bolech, in ``Concepts in Electron Correlation'', edited by A. C. Hewson and V. Zlatic, NATO Science Series II: Mathematics, Physics and Chemistry - Vol. 110 (Kluwer), (2003) 


\bibitem{Glazman88} L. I. Glazman and M. E. Raikh, JETP Lett. {\bf 47}, 452 (1988). 

\bibitem{Oreg03} Yval Oreg and David Goldhaber-Gordon, Phys. Rev. Lett. {\bf 90}, 136602 (2003).

\bibitem{Rau} Ron M. Potok, Ileana G. Rau, Hadas Shtrikman, Y. Ored, and David Goldhaber-Gordon, unpublished.

\bibitem{Gordon98} D. Goldhaber-Gordon, H. Shtrikman, D. Mhalu, D. Abusch-Magder, U. Meirav, and M. A. Kastner, Nature {\bf 391}, 156 (1998).

\bibitem{Cronenwett98} S. M. Cronenwett {\it et al.}, Science {\bf 281}, 540 (1998).

\bibitem{schmid00} J. Schmid, J. Weis, K. Eberl, and K. v. Klitzing,
Phys. Rev. Lett. {\bf 84}, 5824 (2000).


\bibitem{Pustilnik01} M. Pustilnik and L. I. Glazman, Phys. Rev. Lett. {\bf 87}, 216601 (2001).

\bibitem{zarand04} W. Hofstetter and G. Zarand, Phys. Rev. B {\bf 69}, 235301 (2004).

\bibitem{pepin} P. Coleman and C. Pepin, Phys. Rev. {\bf{B 68}},
220405 (2003).

\bibitem{pankaj} P. Mehta, L. Borda, G.Zarand, N. Andrei, and P. Coleman,
Phys. Rev. B {\bf 72}, 014430 (2005).
 
\bibitem{indranil} P. Coleman and I. Paul, Phys. Rev. B {\bf 71}, 035111 (2005).

\bibitem{Anna05} A. Posazhennikova and P. Coleman, Phys. Rev. Lett. {\bf 94}, 036802 (2005).



\bibitem{MW92} Y. Meir, N. S. Wingreen, Phys. Rev. Lett. {\bf 68}, 2512 (1992).


\bibitem{decoherence}G. Zarand, L. Borda, J. von Delft and N.Andrei
Phys. Rev. Lett. 93, 107204 (2004).
\end{thebibliography}
\end{document}